\documentclass[]{article}

\usepackage[a4paper,top=2cm,bottom=2cm,left=3cm,right=3cm,marginparwidth=1.75cm]{geometry}

\usepackage{graphicx}
\usepackage{amsthm}
\usepackage{fancyhdr}
\usepackage{authblk}
\usepackage{wrapfig}
\usepackage{listings}
\usepackage{xcolor}
\usepackage{url}
\usepackage{caption}
\usepackage{subcaption}
\usepackage{acronym}

\usepackage{hyperref}
\hypersetup{
colorlinks = true,
linkcolor = black,
linkbordercolor = {white},
citecolor = blue,
urlcolor=black,
}

\usepackage{booktabs}
\usepackage{adjustbox}
\usepackage{amssymb}% http://ctan.org/pkg/amssymb
\usepackage{pifont}% http://ctan.org/pkg/pifont
\acrodef{TPS}{Transaction Per Second}

\theoremstyle{definition}
\newtheorem{exmp}{Example}[section]

\providecommand{\keywords}[1]
{ \small	
  \textbf{\textit{Keywords---}} #1
}

\title{ \large \bf QPQ 1DLT: A SYSTEM FOR THE RAPID DEPLOYMENT OF SECURE AND EFFICIENT EVM-BASED BLOCKCHAINS}
\begin{document}

\author{\rm Simone Bottoni}
\author{\rm Anwitaman Datta}
\author{\rm Federico Franzoni}
\author{\rm Emanuele Ragnoli}
\author{\rm Roberto Ripamonti}
\author{\rm Christian Rondanini}
\author{\rm Gokhan Sagirlar}
\author{\rm Alberto Trombetta}

\affil{QPQ} 
\affil{\textit {\textit {eragnoli@qpq.io}, {crondanini@qpq.io}, gsagirlar@qpq.io}}

\date{V1.0 \quad  16-08-2022  \quad Zug/London/Dublin  }
\maketitle

\begin{abstract}
Limited scalability and transaction costs are, among others, some of the critical issues that hamper a wider adoption of distributed ledger technologies (DLT). That is particularly true for the Ethereum \cite{wood2014ethereum} blockchain, which, so far, has been the ecosystem with the highest adoption rate. Quite a few solutions, especially on the Ethereum side of things, have been attempted in the last few years. Most of them adopt the approach to offload transactions from the blockchain mainnet, a.k.a. \emph{Level 1} (L1), to a separate network. Such systems are collectively known as \emph{Level 2} (L2) systems. While mitigating the scalability issue, the adoption of L2 introduces additional drawbacks: users have to trust that the L2 system has correctly performed transactions or, conversely, high computational power is required to prove transactions’ correctness. In addition, significant technical knowledge is needed to set up and manage such an L2 system. To tackle such limitations, we propose 1DLT: a novel system that enables rapid and trustless deployment of an Ethereum Virtual Machine based (EVM-based) blockchain that overcomes those drawbacks.

\end{abstract}

\keywords{Blockchain, EVM, Layer Two, Scalability, Network Fees}
\newpage
%\tableofcontents
\newpage

% STORYLINE

% Introduzione - Ethereum L1 scalability state of the art ?
% Layer 2 Limitation - about L2 current limitations
% An overview of 1DLT - our solution overview
% Architecture of 1DLT - our solution architecture
% QPQ Ethereum Node -  our ethereum node architecture
% 	The execution flow - how state update is achieved
% Differences with current Ethereum implementations
% Consensus as a service - CaaS explanation
% Experiments and performance discussion - Experiments 
% Conclusion and future works -  

\section{Introduction} \label{sec:introduction}
The current high demand for Ethereum~\cite{wood2014ethereum} leads to slow transaction throughput (15-30 transactions per second~\cite{ethtps}), expensive gas prices, and poor user experience for the majority of dapps (decentralised apps), Web3 projects, and end users. This limits the potential use cases, like in DeFi (decentralised finance), where high fees and scalability drawbacks enable only entities with vast economic power to trade profitably.\\

A notable and very recent example of extremely high gas prices and network congestion occurred with the launch of a new NFT for the \textit{Bored Ape Yacht Club} metaverse \cite{apes}. Indeed, during the launch, the Ethereum blockchain crashed due to traders outbidding each other by paying higher gas fees in order to execute their transactions at higher speeds. Ethereum users would have spent up to 7,000\$ (2.6 ethers) to mint a 5,846\$ NFT land deed for the virtual world, which in some cases resulted nevertheless in a failed transaction. Likewise, a user trying to send 100\$ in crypto between two wallets would need to pay a fee of 1,700\$. As of July 7 2022, the cost of the above-mentioned NFT floats around 3,000\$.\\
% @christian: inconsistent notation: 23,000 strand for twentythree or twentythree thousand? Also is such number still correct?
% c:we can use the chat, numebr is thousand. the number are related to the old attack right now 2.6 EHT = 3000$

% \tbd{Tom comment in short} 
As it is well known, the Ethereum blockchain requires a massive amount of power for its operations. In fact,
its annual energy consumption is estimated around 112 TWh~\cite{EthEnergyConsumption}, which is comparable to that of Austria.
A single transaction over Ethereum is equivalent to the power consumption of an average US household over 9 days~\cite{ethEnergy}.
However, such huge amount of power does not result in a comparable computational power. In fact, almost all the computational power is used for computing hash values, with the power left for other operations being just a tiny fraction of the computational power of a Raspberry Pi 4~\cite{web3fraud}, see Section \ref{sec:FCEdiscussion} for more details.
%the current expensive transaction fees and over the 99\% of the computational power wasted in hashing functions make a single-board %computer like the Raspberry Pi 4 have 5000 more compute power than Ethereum ~\cite{web3fraud}.
\\

Therefore, scaling solutions become crucial to increase network capacity in terms of speed and throughput. However, improvements to scalability should not be at the expense of decentralisation or with the introduction of a trusted third party. Traditionally, scalability solutions are based on off-chain systems, collectively known as ``Layer 2" (L2). L2 solutions are implemented separately from the ``Layer 1" (L1) Ethereum mainnet and do not require changes to the existing Ethereum protocol. In L2 solutions, transactions are submitted to the nodes of the L2 system instead of directly to L1 nodes. Thus, L2 solutions handle transactions outside the Ethereum mainnet and take advantage of the architectural features of the mainnet to allow high decentralisation and security. The existing L2 systems show a wide array of trade-offs among critical aspects like throughput, energy consumption, security guarantees, scalability, gas fees, and loss of trustlessness.\\

In this work, we present \emph{One DLT} (1DLT), a novel, modular system for the rapid deployment of EVM-based blockchains, that avoids the pitfalls of many of the existing L2 solutions. The rest of this paper is organised as follows: Section \ref{sec:Limitation} reviews the trade offs of current solutions; sections \ref{sec:1CLDT}, \ref{sec:architecture}, and \ref{sec:EthereumNode} describe our system; Section \ref{sec:caas} describes the Consensus-as-a-Service mechanism at the core of 1DLT; Section \ref{sec:Bridge} show the 1DLT Bridge architecture; Section \ref{sec:experiments} exhibits a set of preliminary experimental results; Finally Section \ref{sec:conclusion} concludes the work and describes our next steps.
\\

\section{Layer 2 limitations } \label{sec:Limitation}
There are several solutions available in the L2 ecosystem~\cite{L2eth} (e.g., Optimistic Rollups~\cite{optrollup}, ZK-rollups~\cite{zkrollups}, State channels~\cite{statechannels}, Sidechains~\cite{sidechains}), with a lot of variation in terms of advantages and limitations. In the following, we list the most fundamental limitations, and we correlate them to some of the solutions adopted by the Ethereum ecosystem: 

\begin{itemize}
    \item 	\textbf{Limited expressive power}: some solutions do not support EVMs (e.g., several ZK-rollups, Plasma~\cite{plasma}, Validium~\cite{validium});  other solutions support application-specific computations and require specialised languages (e.g., StarkWare's Cairo~\cite{StarkwareCairo});
    \item	\textbf{L2 nodes}: some solutions  use operators and validators that can influence transaction ordering, reversing the principle of trustlessness that might lead to abuses or frauds (e.g., Optimistic Rollups, Sidechains);
    \item	\textbf{Liveness requirement}: some solutions need to periodically watch the network or delegate this responsibility to someone else to ensure security (e.g., Plasma);
    \item	\textbf{High computational power to compute proofs}: some solutions require high computational power to compute proofs, which can be too expensive for dapps with little on-chain activity (e.g., ZK-rollups, Validium);
    \item	\textbf{Reduced decentralisation}: some solutions adopt centralised methods to mediate the implementation of weak security schemes  (e.g., Sidechains);
    % (e.g., ZK-Rollups setup promotes a centralised scheme) . Some security scheme assumes a level of unverifiable trust 
    \item	\textbf{Limited throughput}: some solutions claim to theoretically achieve high  transactions per second (tps) but are practically limited in their implementations (e.g, StarkWare \cite{Starkware} theoretically achieves 2,000 tps, whereas the actual limit in real-world deployments is 650 tps);
    \item	\textbf{Not L2}: some solutions cannot technically be considered as L2 since they use separate consensus mechanisms that are not secured by the respective L1 (e.g., Sidechains). As such, these solutions cannot inherit from the L1 its security guarantees (e.g., resilience against chain tampering for Ethereum);   
    % \tbd{1. need to explain/elaborate, so what? it thus compromises on security? Also, isn't that also true for 1DLT?}
    \item	\textbf{Private channels}: some solutions  implement private channels, which is not a viable solution for infrequent transactions (e.g., State channels);
    \item	\textbf{Long on-chain wait times}: some solutions require long wait times for on-chain transactions due to potential fraud challenges (e.g., Optimistic Rollups, Validium);
    \item	\textbf{Data availability}: some solutions generate proofs that require off-chain data to be always available (e.g., Validium).
    % \item   \textbf{High energy consumption}:
    % \item   \textbf{High computational cost}:

    %\item	Vulnerability to Quantum computing attacks due to the use of  non-quantum resistant cryptographic algorithms.
\end{itemize}

While the list above is by no means exhaustive (indeed, the L2 landscape is so dynamic that novel solutions, prototypes and products are being introduced to the market frequently), it is indicative of how, while there clearly exist attempts at overcoming those limitations, there is not a single solution that can fix them all. It is important to note that a consequence of some of the limitations is the generation of security risks. Indeed, chains with relatively small ecosystems that provide consensus can lead to fallacies of abuse and fraud. Attackers, or the node maintainers themselves, may tamper with blockchain data ordering or validation to take advantage, such as redirect funds, perform flash loans or double-spending attacks, etc.\\

Last but not least, in most of the solutions above, users aiming at creating a private or public Ethereum network must rely on Ethereum \emph{clients} (also known as \emph{implementations})\footnote{A client is an implementation of Ethereum that verifies all transactions in each block, keeping the network secure and the data accurate~\cite{ClientsNode}} like Geth~\cite{Geth} and Erigon~\cite{Erigon}. This approach requires the user to have a significant technical knowledge and resources to maintain the nodes, with all the related costs and requirements of technical know-how. 
% \newline

%Our goal is to overcome such limitations. Thus, in what follows we explore the approaches and choices we adopted in order to cope with them.
% With regard to this list, our proposed solution aims to overcome or simplify all these limitation, thanks to approaches, such as , which we details 
% simplify and thus offer a solution that can overcome this limitation or simplify them, with different approaches, in the next section we explore the proposed solution and how and why we can overcome these limitation.
% the proposed architecture simplifies and thus offers a solution to this drawback

\section{An overview of 1DLT}\label{sec:1CLDT}

This work has been inspired by the user experience of Cloud Service Providers (CSP) and Web-based applications, guided by the principles of DLTs. Indeed, while CSP dashboards might not be perfect, a user is directed via graphical interfaces and dashboards through the process that enables to complete the setup, configuration, billing, and management of the chosen service. The interaction with such an interface is performed without any need of deep knowledge of the underlying technologies.\\
\\
Hence, our goal is to provide a system that:
\begin{itemize}
\item has a very low technical barrier of entry that streamlines and simplifies the deployment of a (public/private) EVM-based blockchain, as customarily happens for web-based services and CSP dashboards, without discarding the programmability of the EVM;
\item maintains the security and trustlessness of DLTs while improving scalability and lowering gas fees;
\item removes the risks associated with the L2 governance and fraud or abuse detection.
\end{itemize}

This is achieved with a modular, multilayered cloud-native architecture (see Section \ref{sec:architecture}) that decouples the transaction layer from the consensus layer. Thus:

\begin{itemize}
\item  1DLT connects to different blockchains in order to leverage their consensus mechanisms. We refer to this as \emph{Consensus-as-a-Service}  (CaaS)  (see Section \ref{sec:caas} for further details);

\item 1DLT removes the risks associated with the L2 governance and fraud detection. Indeed, all transactions processed by 1DLT are sent to the L1 public blockchain, which is used as a consensus resource. resource. This also allows to inherit the security guarantees of the L1 public blockchain that provides consensus;

\item 1DLT does not suffer from long wait times used in L2 to detect and avoid frauds. Fraud detection can be performed by checking receipts and confirmation messages of the public blockchain used as the consensus provider and local transactions' meta-data sent by CaaS;

\item 1DLT is EVM-based, and it directly supports smart contracts written in the Ethereum programming language, Solidity, and does not require the adoption of L2 specific languages, such as Cairo\footnote{https://starkware.co/cairo/};

\item 1DLT maintains trustlessness. Since every transaction is sent to a public blockchain via CaaS, 1DLT is as decentralised as the blockchain to which CaaS connects to;

\item 1DLT transaction throughput is limited by the throughput of the public blockchain that it connects to via CaaS. 
% However, since there are public blockchains that are better performing and scalable than Ethereum, this improves scalability;
Therefore, 1DLT outperforms Ethereum by connecting to blockchains with higher throughput and better scalability.
Additionally, 1DLT performance can be improved by connecting to different blockchains over time based on load on a given network, and furthermore, the modular architecture makes it ready to leverage on new, faster blockchain networks as and when they come into being;

% Also, it needs to be noted that the throughput of 1DLT can be significantly increased by connecting to the optimal consensus provider, that is enabling to manage and select more than one Consensus providers simultaneously. 
 %Also, it needs to be noted that the throughput of 1DLT can be significantly increased by connecting and optimally managing more that one Consensus providers simultaneously. 

%\tbd{1. What are the cost implications? 2. If multiple public blockchains are used in parallel, what is the security implication claimed in point 2 above? I am unsure whether these need to be addressed here (1 probably should be), but they need to be addressed somewhere in the document.}

\item 1DLT allows to significantly reduce the transaction fees required to perform operations like payments, smart contract deployments, and token swaps, thanks to CaaS.   

\end{itemize}

The following example is a snippet of the user experience with 1DLT.

\begin{exmp}\label{ex:1}
%  Use case example here 
Due to the high transaction fees and long transaction confirmation times, Alice wishes to move her dapp performing \textit{NFTs Auctions} from the Ethereum mainnet to another blockchain, that is 1DLT. However, she does not want to change her codebase. To achieve that, Alice deploys a small, private EVM-based blockchain with 1DLT, formed by a pair of nodes.

(i) Alice registers with the QPQ authentication system\footnote{A user must be authenticated to perform any action, thus a trusted authentication system is entitled to handle the user registration and management. Note that this trusted entity, while serving as a gateway for participants, is not relied upon for accountability of the actions of the participants. With our approach, the latter is achieved in a trustless manner using CaaS.} and receives a redemption code for 1DLT blockchain creation;

(ii) Alice specifies the blockchain name, description, token name, and token symbol. Then, with the redeemed code, she creates the first QPQ Ethereum node of the private blockchain, choosing configuration parameters such as cloud provider, virtual machine, hostname, etc.;

(iii) after specifying blockchain and nodes' parameters, Alice waits a short period for the execution of the setup procedure to start the deployment of her dapp;

(iv) Alice is now ready to deploy her dapp using the same Web3 API and process that she used in Ethereum, which in this case consists in sending a deployment transaction through the Web3 API; 

(v) finally, upon receiving the deployment confirmation within instants, she and her customers are ready to interact with the dapp.
\end{exmp}
\section{Architecture of 1DLT}\label{sec:architecture}
The architecture of 1DLT is based on a modular approach, and it has two main components: a private EVM-based node (called \emph{QPQ Ethereum Node}) and Consensus-as-a-Service (CaaS), a module that connects to external consensus providers. The CaaS module allows the QPQ Ethereum Node to access an external DLT and leverage its consensus mechanism. Therefore, the 1DLT architecture abstracts and decouples the transaction layer from the consensus layer.\\
%Indeed, this generates a consensus on-demand method that exploits the Consensus protocols and mechanisms of external, public blockchains.

\begin{figure}[!htbp]
\centering
\includegraphics[width=0.75\textwidth, keepaspectratio]{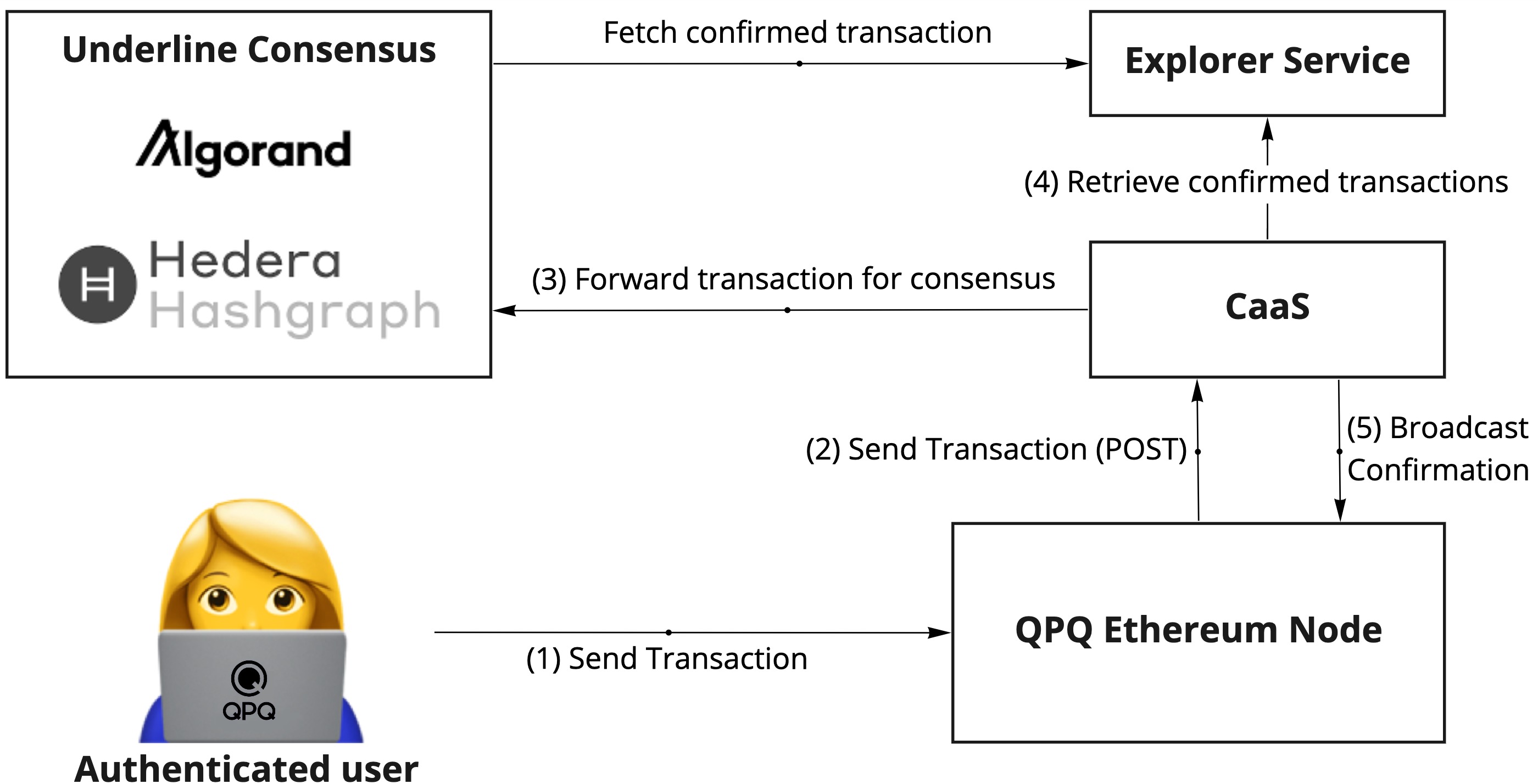}
\caption{QPQ 1DLT architecture }\label{fig:architectureHighlevel}
\end{figure}

% 
A high-level overview of the components of 1DLT and their interactions is shown in Figure \ref{fig:architectureHighlevel}.
%, and user.

\begin{exmp}\label{ex:2}
We continue with the scenario introduced in Example \ref{ex:1}, in which Alice did the setup of a node, and interacts with it by deploying a smart contract and sending a transaction. We expand on the operation flow sketched in Figure \ref{fig:architectureHighlevel}:

(i) Alice calls the Web3 API provided by the QPQ Ethereum node to send a transaction, i.e. $eth\_send\_\\raw\_transaction$;
% or the $deploy$ function for the smart contract creation.

(ii) the QPQ Ethereum node receives the submitted transaction and forwards it to the Consensus-as-a-Service (CaaS) handler using a POST message. The QPQ Ethereum node leverages CaaS to achieve consensus for the transaction. 
We reiterate that CaaS may operate with different consensus protocols, and we refer to Section \ref{sec:caas} for the explanation of how CaaS chooses a DLT for transaction dispatching;
%By doing so, we offload all the consensus complexity off-chain, meaning that we can drastically change the node design to reduce the computational efforts required inside the network (see section \ref{sec:differences} for details). As such, we enable the QPQ Ethereum node to achieve better performances, such as lower transaction costs and finality.
% This is due to the fact that we do not have to reach the consensus inside our network, and so we do not need any computation effort

(iii) CaaS forwards the transaction to the chosen DLT to achieve consensus;

(iv) then, once the transaction is confirmed, CaaS retrieves it using a blockchain explorer service (e.g., Hedera mirror service~\cite{Hederamirrorservice});
% and

(v)  finally, CaaS broadcasts the confirmation of the transaction to the QPQ Ethereum node, such that it can update its state. 
% \tbd{need to include a short explanation for the situation when a transaction is not confirmed}
\end{exmp}

\section{QPQ Ethereum Node} \label{sec:EthereumNode}
While there are well-known and widely used implementations of Ethereum nodes, notably Geth and Erigon, in order to overcome some of the limitations of Section \ref{sec:Limitation}, we engineered our own Ethereum node with a much simpler architecture (see Figure \ref{fig:node architecture}). Section \ref{sec:differences} describes the differences with a standard implementation. Generally speaking, an Ethereum node contains the following components: a Web3 API, a state handling mechanism (that is, the set of tries storing the information composing the state~\cite{Trie}), a database, an Ethereum Virtual Machine (EVM)~\cite{EVM}, a p2p network, a transaction pool, and a consensus protocol. In the following, we present a detailed description of the modules of our proposed architecture:

\begin{figure}[!htbp]
\centering
\includegraphics[width=0.60\textwidth, keepaspectratio]{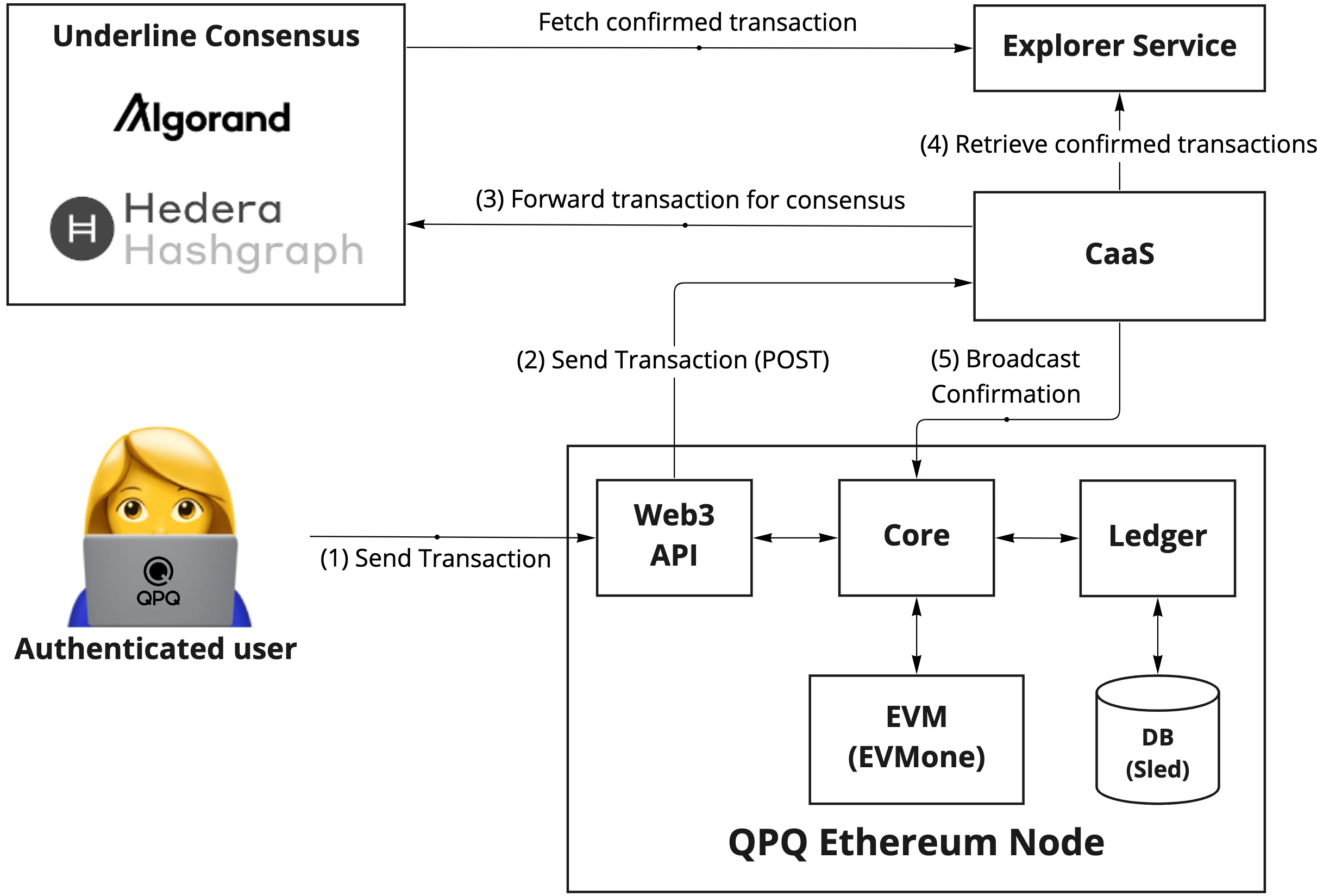}
\caption{QPQ Ethereum node architecture  }\label{fig:node architecture}
\end{figure}

\begin{itemize}
    \item \textbf{EVM module}:  it is a sandboxed virtual stack machine. The purpose of the EVM module is to compute the new system state by executing an instruction specified in a smart contract.
    In order to connect the node to a local, private EVM, the Ethereum Client-VM Connector API (EVMC) is used, as shown in Figure \ref{fig:evmc}. The EVMC is the low-level interface between Ethereum Virtual Machines (EVMs) and Ethereum clients, which -- on the EVM side -- supports classic EVM1\footnote{Ethereum 1.x is a codename for a comprehensive set of upgrades to the Ethereum mainnet intended for near-term adoption.} and ewasm\footnote{Ethereum flavoured WebAssembly is a subset of the WebAssembly (wasm) format used for contracts in Ethereum.}. On the client side, EVMC defines the interface for accessing the Ethereum environment and state.
    A very relevant feature of EVMC lies in the fact that nodes can connect with other non-Solidity based virtual machines.
    
    % references 
    % https://arxiv.org/pdf/2012.01032.pdf
    % https://evmc.ethereum.org/
    % https://github.com/ethereum/evmc
    % https://blog.secondstate.io/post/20191025-compile-and-deploy-an-erc20-contract-on-ewasm/
    % https://docs.ethhub.io/ethereum-roadmap/ethereum-1.x/
    \begin{figure}[!htbp]
    \centering
    \includegraphics[width=0.50\textwidth, keepaspectratio]{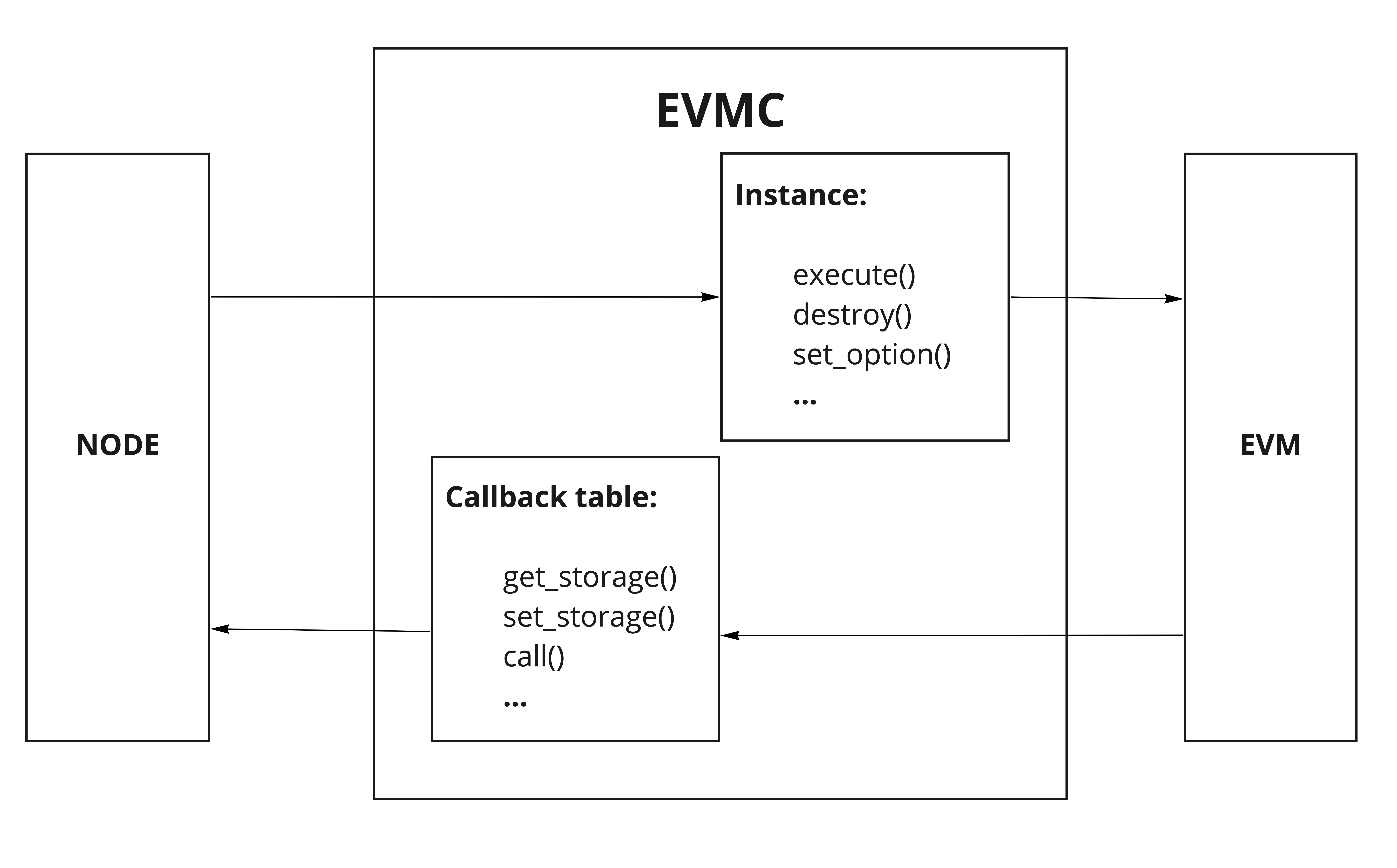}
    \caption{EVMC API}\label{fig:evmc}
    \end{figure}
    % references 
    % https://fisco-bcos-documentation.readthedocs.io/en/latest/docs/design/virtual_machine/evm.html or
    % https://chowdera.com/2021/07/20210725192100089H.html
    % https://moralis.io/evm-explained-what-is-ethereum-virtual-machine/
     
    1DLT deploys a standalone C++ EVM implementation, called \emph{EVMone}~\cite{EVMone}. The EVMone EVM can be imported as a module by an Ethereum client and provides efficient execution of smart contracts written in an EVM-compliant scripting language. 
    % The reason we opted evmone are for performace and to be compliant with the yellow paper -~ have to explain why

    \item {\bf Web3 API module}: it handles incoming transactions and the communication with the CaaS handler. This module mostly works in the same way as in an Ethereum implementation, except for a customization that allows it to interact with CaaS. The module exposes a Web3-compatible API supporting modern Ethereum development tools and wallets (e.g., Metamask~\cite{metamask}, Hardhat~\cite{hardhat} and Web3.js~\cite{web3APIjs}). 
    As an example, the API supports $eth\_send\_raw\_transaction$ the same way as other Ethereum implementations do. On the other hand, it does not support mining-related methods like $eth\_isMining$, since we do not have nor need a consensus mechanism that involves mining.

     \item {\bf Ledger and state transitions module}: regarding the ledger used for storing transactions' results, we opted for an Ethereum-compatible persistent ledger implementation using Merkle Patricia Trees (also known as (Merkle) Tries). This enables us to rely on the same storage structure of a standard Ethereum node (that is, a State Trie, Receipt Trie, Transaction Trie, and Storage Trie~\cite{wood2014ethereum}). 
        % https://medium.com/@eiki1212/ethereum-state-trie-architecture-explained-a30237009d4e
    % What differentiates our solution from an Ethereum client is that we do not have a list of tries but only one for the ledger.

    To this end, instead of the LevelDB DBMS~\cite{leveldb} used in other Ethereum implementations, we opted for Sled~\cite{Sled}, an embedded key-value store written in the Rust programming language.  Sled provides atomic single-key operations, including compare and swap operations. It is designed to be used as a construction component to build larger stateful systems. Sled is optimised for modern hardware. It uses lock-free data structures to improve scalability and organises storage on disk in a log-structured way optimised for SSDs.
    %  sled, which is a high-performance embedded database with an API that is similar to a BTreeMap<[u8], [u8]>, but with several additional capabilities for assisting creators of stateful systems.
    % https://github.com/spacejam/sled
    % https://docs.rs/sled/latest/sled/#:~:text=sled%20is%20a%20high%2Dperformance,and%20all%20operations%20are%20atomic.
    % https://dbdb.io/db/sled

    We do not perform complex operations to achieve state transitions, such as the staged sync~\cite{Erigonstage} in Erigon, as well as for block cutting. 
    As previously discussed, CaaS validates the transactions using a public consensus resource, and this enables us to update the ledger state (EVM execution) and the block cutting directly without needing the consensus on the operation again.
    This enables us to perform the block cutting in multiple ways, without any constraints. To this end, we opted to cut the block every $\Delta$ seconds (e.g., 10 seconds), after checking that the block is not empty.

    \item {\bf Core module}: The \emph{Core} module manages and coordinates the interaction of the QPQ Ethereum Nodes' modules. It retrieves consensus updates of the processed transactions from the Consensus as a Service module (see section \ref{sec:caas}) to perform state changes.
    Unlike other Ethereum node implementations (e.g. Erigon) that perform complex operations for state updates (e.g. Erigon performs ``stages sync''), a QPQ Ethereum Node can directly update the state after retrieving the consensus updates of transactions from CaaS.
    %Thus, QPQ Ethereum Nodes' core modules can be considered as lightweight compared to other Ethereum node implementations.

\end{itemize}

\subsection{ The execution flow } \label{sec:executionFlow}
\begin{figure}[!htbp]
\centering
\includegraphics[width=0.70\textwidth, keepaspectratio]{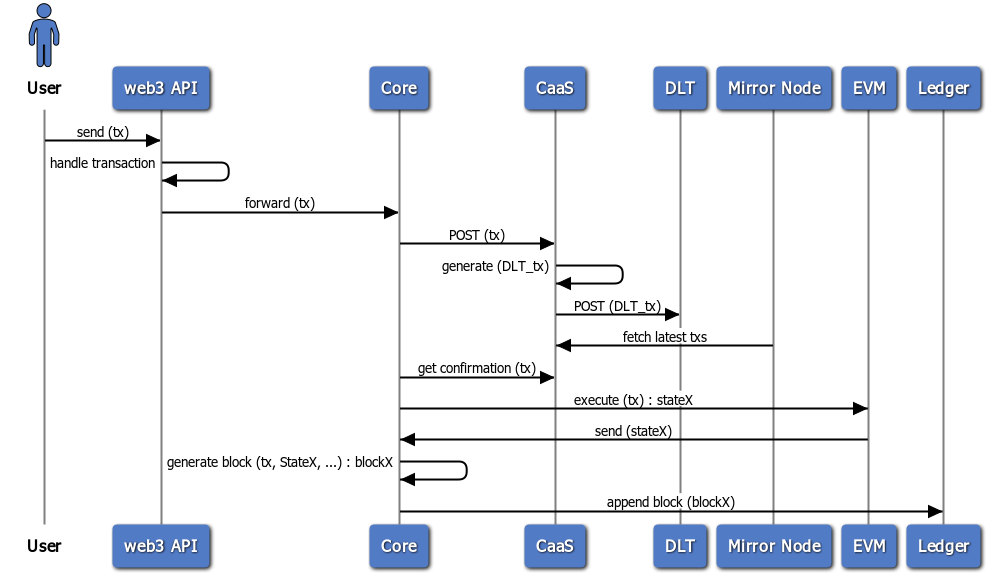}
\caption{ Sequence Diagram of a state update in a QPQ Ethereum node}\label{fig:sequence}
\end{figure}

In what follows, we briefly describe the steps needed to perform a state update in a QPQ Ethereum node upon receiving a transaction (see the sequence diagram in Figure \ref{fig:sequence}).
\begin{enumerate}

% \item  A user has to register with the platform ( QUO authentication )
% \item  A user exploiting the UX interface sends a transaction to the QPQ Ethereum node. 
\item 	The transaction is sent to the QPQ Ethereum node through the Web3 API;
\item 	the Web3 API module handles the transaction and sends a POST request to CaaS with the transaction wrapped inside the data field;
\item 	CaaS handles the transaction and connects to one of the chosen blockchains (e.g., Hedera) to reach consensus on the transaction; 
\item 	CaaS communicates with an external service  (e.g. a Hedera mirror node) to retrieve the transaction confirmation;
\item   CaaS then broadcasts the confirmation message (i.e., the hash of the transaction with the consensus proof) of the transaction to the QPQ Ethereum node;
\item 	the EVM of the node executes the transaction, updating the state;
\item   a block is then created if $\Delta$ seconds are passed (the time interval is a configurable parameter whose default is set to 10 seconds);
\item 	finally, the state transition result is stored in the Sled database.
\end{enumerate}

\subsection{ Differences with current Ethereum implementations} \label{sec:differences}

Our approach demands that more than one protocol may be available for transaction consensus. As such, several relevant changes have been introduced that differentiate the node implementation used in 1DLT from the standard one:

\begin{itemize}
    \item {\bf External Consensus}: in general, a consensus protocol has to be included in the internal architecture of an Ethereum node -- like PoW for a mainnet, or PoA for a testnet node. Instead, we do not rely on an internal module, and delegate consensus to CaaS. This enables us to have the same level of liveness and safety guarantees of the chosen DLT while decreasing the overall complexity of the node.

    \item {\bf No transaction pools}: in general, in an Ethereum node the transactions waiting for confirmation by means of an internal consensus mechanism are placed into a transaction pool. Since our solution relies on an external consensus via CaaS, all newly arrived transactions are forwarded directly to CaaS for confirmation without putting them in a queue.
    % Then, CaaS retrives the confirmed transaction using a blockchain explorer service (e.g., Hedera mirror service) and broadcasts the confirmation of the transaction to the QPQ node who sent the transaction and to all other QPQ nodes in the network.  The block creation perfomed by
    % CaaS communicates with each node regarding the transaction order, enabling block creation without further delay.
    The benefits from this choice are a significantly simplified design and overall increased performance, as shown in section \ref{sec:experiments}.
    
    \item {\bf Lightweight Core module}: as already mentioned in the previous section, the Core module is thoroughly simplified, since there is no transaction pool to manage, and it does not have complex state transitions (e.g., staged sync) as the consensus is retrieved from CaaS.
    
    %\item {\bf No peer-to-peer}.  Usually, the peer-to-peer module oversees synchronisation between other peers in the network. However, 1DLT does not have a module in charge of the p2p communication. Instead, we rely on CaaS to achieve the synchronization by broadcasting to each node in the network a newly confirmed transaction. In effect, the CaaS substitutes the p2p module.

    \item {\bf Difference in the Web3 support}: having an external consensus provider implies that several methods are not supported by our implementation. In particular, methods related to mining (e.g., $eth\_getMining$, $eth\_submitHashrate$, $eth\_coinbase$), to uncles (e.g., $eth\_submitHashrate$), and to Ethereum protocol (e.g., $eth\_protocolVersion$) are not supported. 
    % 
    % and to account related (e.g., $eth\_sendtransaction$, $eth\_getAccounts$)
    % eth_signTransaction / send transaction
    
\end{itemize}
\section{Consensus-as-a-Service } \label{sec:caas}

Consensus-as-a-Service (CaaS) is the module that allows a QPQ Ethereum node to access an external, public consensus protocol with an on-demand approach. Its key feature is the introduction of an abstract layer that enables the access to different DLTs through a single, uniform interface.
% a layer that uniformly presents different DLTs, abstracting their details and defining a single interface to access their services. 
Such a layer allows 1DLT networks' nodes to offload consensus complexity by enabling them to achieve better performance, such as higher throughput and faster transaction finality with low transaction costs. Moreover, relying on an external consensus provider enables 1DLT to inherit the security model of the chosen DLT. 
% 
% 
% Last but not least, we remark that the use of CaaS as a trusted party is mitigated by having all processed transactions public and auditable.
% Last but not least, we remark that CaaS overcomes any issue that may arise from the presence of a trusted third party, as transaction handled by CaaS are public and easily auditable.
Last but not least, we remark that the CaaS approach eliminates any issue that may arise from the presence of a trusted third party, since transactions are public and easily auditable.
% As the user can audit the external DLT, but caas can still modify it % minimize the issue of having a trusted third-party
% practically speaking, we're still trusting caas to handle transactions correctly (i.e., to not tamper with them, and to forward all of them to the dlt).the fact that we can audit what transactions are actually processed is a plus, which mitigates but do not remove the trust in caas
% 

% Write that it can still be modified, however, in that case it wouldn't be valid anymore and so cannot be executed
While CaaS could attempt to tamper with handled transactions, however, it would make the transaction proof invalid, preventing the transaction execution.  
In fact, CaaS cannot tamper with handled transactions in any way, as each transaction proof required for an audit process can be retrieved from the target DLT.\\
\\
After having chosen a suitable DLT, CaaS interacts with it by creating
a channel, that is used to publish messages that contain transactions' information using the CaaS's DLT interface. 
The exchanged messages are stored in a time-series database (in the current implementation, we use timescale v2.6.0-pg1) to guarantee benefits over traditional Relational database management systems (RDBMS) or vanilla PostgreSQL; these benefits include time-oriented features, higher data ingest rates and query performance.
% the delivery liveness of the network in an efficient way.
% 
% Much higher data ingest rates, especially at larger database sizes.
% Query performance ranging from equivalent to orders of magnitude greater.
% Time-oriented features.
% 
% 
Additionally, each delivered message comes with the receipt of the transaction from the chosen DLT (e.g., an Hedera transaction receipt), which is an auditable proof that the transaction has been correctly processed by the DLT itself.\\

% \tbd{Paralellization and multi instances}
CaaS manages communication channels, I/O operations, and DB operations concurrently. In essence, CaaS spawns and manages multiple threads dedicated to the  message exchange of different 1DLT networks. In particular, each 1DLT network has at least one dedicated communication channel managed by CaaS, allowing high message processing throughput with low latency. 
Having dedicated communication channels also allows multiple 1DLT networks to co-exist while isolating them from each other.
% , thus preserving the privacy of the transactions processed by all 1DLT networks.

\newpage
\section{Bridge} \label{sec:Bridge}

Blockchains are siloed environments that cannot communicate with each other, as each network has its own protocols, native assets, data, and consensus mechanisms.
Blockchain bridges, or cross-chain bridges \cite{ethBridgeIntro}\cite{ethBridgeIntro2}, are a possible solution for enabling interoperability between different blockchains.
% 
% from a theoretical point of view, blockchain A communicate with blockchain B simply by exchanging message. 
% In reality, due to trust boundaries between the two blockchain, they cannot simply talk to each other, 
%In reality, there is no way for blockchain B to reply on the same channel used by A to confirm that the message has been received, due to trust boundaries between the two blockchains.
%Something in the middle is needed to bridge the gap and overcome the trust boundaries. 
%To this end, bridges use different off-chain mechanisms, or actors, that play the role of verifiers acting as the ‘man-in-the-middle’ between blockchains.
% Something in the middle is needed to bridge the gap and overcome the trust boundaries. 
% How a bridge works is mainly differentiated by the role of verifiers, that is, if the bridge use a trusted or trustless system of verifiers.  
%Usually, a trusted or trustless system of verifiers distinguishes how a bridge operates.
% In  \cite{https://blog.connext.network/the-interoperability-trilemma-657c2cf69f17} Arjun Bhuptani, classifies bridges based on how they are verified into natively, externally, and locally verified system.
% 
The interoperability trilemma~\cite{InteroperabilityTrilemma} allows for different bridge designs, for which, a non-standard classification can be based on~\cite{bridgeClassification}:
\begin{itemize}
    \item {\bf Trust model - How they work}: that is, the type of authority used to synchronise the operations. The bridge is referred to as a \textquotedblleft trusted bridge" if there is a central-trusted authority (e.g., Binance bridge~\cite{binancebridge}).
    If not, smart contracts make the bridge a \textquotedblleft trustless bridge" by doing away with the necessity for a reliable third party  (e.g., Connext~\cite{connext}, Hop~\cite{hop}, and other bridges with a simple atomic swap mechanism).
    % If it is a central-trusted authority, then the bridge is classified as a "trusted bridge" (e.g., $Binance <> Ethereum$ bridge ~\cite{binancebridge}). Else, the bridge is a "trustless bridge", as the role of a trusted third party is removed through smart contracts. (e.g., Connext ~\cite{connext}, Hop~\cite{hop}, and other bridges with a simple atomic swap mechanism).
        
    \item{\bf Validation - Validator or oracle based bridges}: that is, the type of mechanism the bridge relies on to validate cross-chain transfers, such as external validator or oracles. %These bridges rely on an external validator set or oracles to validate cross-chain transfers. Examples: Multichain and Across.

    \item{\bf Level - What they connect to}: that is, the type of systems it connects to. If the connection is between blockchains or between a blockchain and an $L2$ system.
    %   \subitem{\bf Native bridges}:  – These bridges are typically built to bootstrap liquidity on a particular blockchain, making it easier for users to move funds to the ecosystem. For example, the Arbitrum Bridge is built to make it convenient for users to bridge from Ethereum Mainnet to Arbitrum. Other such bridges include Polygon PoS Bridge, Optimism Gateway, etc.
   
    \item{\bf Sync mechanism - How they move assets}: that is, the type of mechanism used to transfer assets between blockchains, such as \emph{Lock and mint}, \emph{Burn and mint}, or \emph{Atomic swaps}.
        % There are 3 option,
        % Lock and mint. The bridge lock the asset on chain A and mint asset on chain B 
        % Burn and mint. The bridge brun the asset on chain A and mint asset on chain B 
        % Atomic swap. The bridge swap the asset form chain A to chain B. it is relevant to note that  they rely on smart contracts for assets swap and are generally more trustless (decentralized) and remove the need for a trusted party.

    \item{\bf Functionality - Their function}: that is, the (more or less) specialized interoperability task they are meant for, such as Chain-To-Chain Bridges, Multi-Chain Bridges, Specialized Bridges, Wrapped Asset Bridges, Data Specific Bridges, dapps Specific Bridges, and Sidechain Bridges.
            % \subitem{\bf Generalized message passing bridges}:  – These bridges can transfer assets, along with messages and arbitrary data across chains. Examples: Nomad and LayerZero.

            %  \subitem{\bf Liquidity networks}: These bridges primarily focus on transferring assets from one chain to another via atomic swaps. Generally, they don’t support cross-chain message passing. Examples: Connext and Hop.

\end{itemize}

\begin{figure}[!htbp] \centering
\includegraphics[width=0.90\textwidth, keepaspectratio]{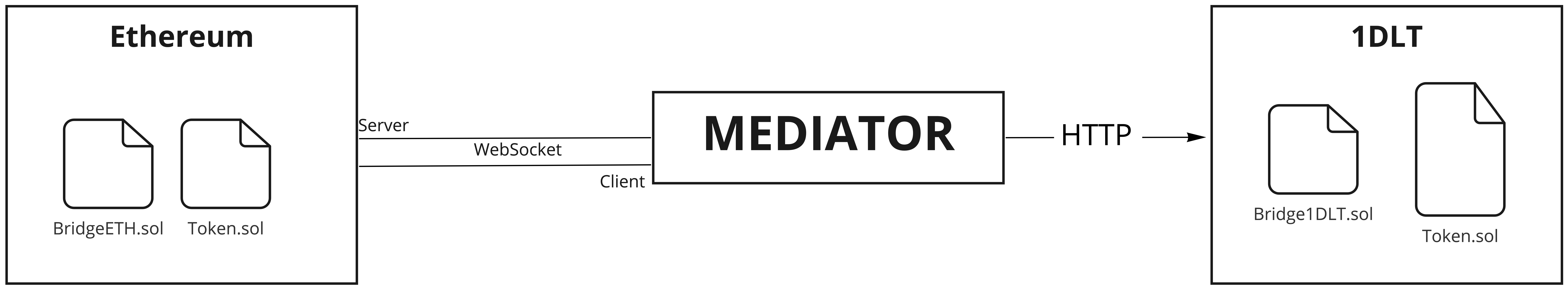}
\caption{ 1DLT bridge architecture}\label{fig:bridgeOverview}
\end{figure}

% 1DLT small intro
% 1DLT features
%  trustless classification
%  lock and mint mechanism 
%  support every evm compatible blockchain not only L1 or L2
%   multi feature -- > has multiple function, this can be tricy to insert
1DLT offers a unique solution for bridges. We enable users to deploy the bridge by providing them with all the tools and components (e.g., smart contracts).
% We cover the back-end while leaving the front-end to the user. 
% 
Since the bridge is operated via smart contracts, which serve as a trusted party, the 1DLT bridge belongs to the set of trustless bridges.
It allows for the bi-directional transfer of ERC20 and ERC721 tokens between 1DLT nodes and EVM-compatible blockchains.
% , through a \emph{Lock and mint} mechanism.
1DLT uses a \emph{Lock and mint} mechanism: on the origin chain (e.g., Ethereum), a lock over the asset is performed, while on the destination chain (e.g., 1DLT), a mint is performed.
Figure \ref{fig:bridgeOverview} provides a high-level breakdown of the bridge's core elements and how they interact.
% The bridge smart contracts are deployed on each target blockchain and their interaction is coordinated via a bridge API, exploiting HTTP and WebSocket. connections. 
Essentially, the bridge is composed of a set of smart contracts, \emph{Bridge.sol} and \emph{Token.sol}, that are deployed on both the source and destination blockchain.
Their interaction is coordinated via a bridge API, called \emph{Mediator}, using HTTP and WebSocket. 
The \emph{Bridge} smart contract the implementation differ, as on the destination chain (i.e., \emph{Bridge1DLT} for 1DLT) the contract design is for burn and mint, while on the origin chain is lock and withdraw (i.e., \emph{BridgeETH}) for Ethereum).
For \emph{Token} the implementation is the same for both the chains.

\begin{figure}[h]
  \centering
\includegraphics[width=0.65\textwidth, keepaspectratio]{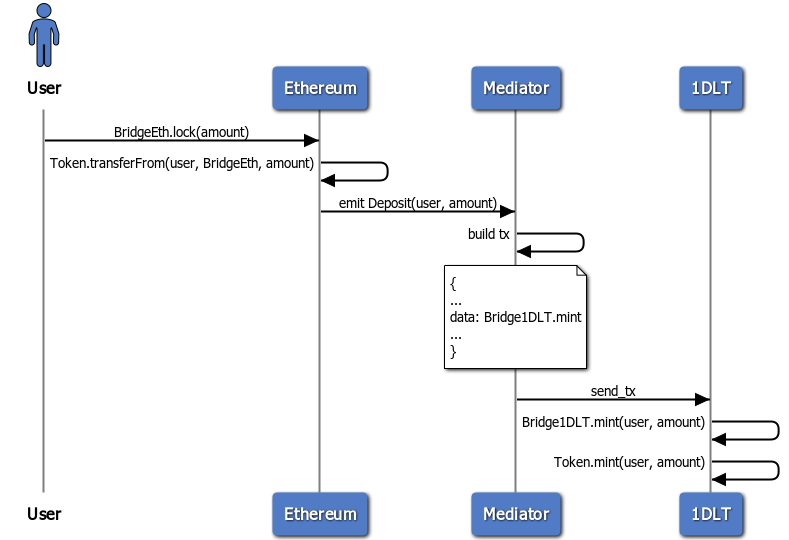}
\caption{ Deposit of ERC20 tokens from Ethereum to 1DLT}\label{fig:onboard}
\end{figure}

% \subsection{ Execution flow }
% how it works under the hood
In what follows, we briefly describe the steps needed to perform the deposit of some ERC20 tokens from  Ethereum to 1DLT (see Figure \ref{fig:onboard}).
The process relies on locking the asset on the Source blockchain, and then in the destination blockchain mint the amount\footnote{To prevent the user from minting an arbitrary amount of token, only the bridge smart contract is entitled to call the mint method in the token smart contract.}. 

\begin{enumerate}
   \item The user sends a transaction to Ethereum, calling the \emph{lock} method defined in the \emph{BridgeEth} smart contract;
   \item The transaction locks the tokens on Ethereum, transferring them to the \emph{BridgeEth} address;
   \item The \emph{BridgeEth} emits a custom \emph{Deposit} event with the address of the receiver on 1DLT and the amount;
   \item The \emph{Mediator} detects the event and retrieves the information;
%   by exploiting the Websocket connection to Ethereum;
   \item The \emph{Mediator} builds a transaction to call the \emph{mint} method defined in \emph{Bridge1DLT} with the event information as parameters;
   \item The \emph{Mediator} sends the new transaction to 1DLT;
   \item 1DLT executes the transaction, which calls the \emph{mint} method of \emph{Bridge1DLT};
   \item The method calls the \emph{mint} defined in \emph{Token};
\end{enumerate}

\begin{figure}[h]
\centering
\includegraphics[width=0.65\textwidth, keepaspectratio]{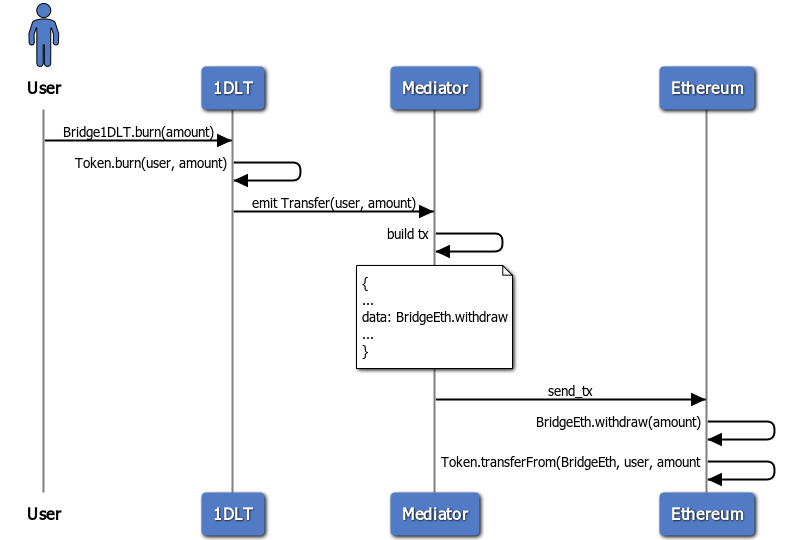}
\caption{ Withdraw of ERC20 tokens from 1DLT to Ethereum}\label{fig:offboard}
\end{figure}

In what follows, we briefly describe the steps needed to perform the withdrawal of some ERC20 tokens from 1DLT to Ethereum (see Figure \ref{fig:offboard}).
\begin{enumerate}
    \item The user sends a transaction to 1DLT calling the \emph{burn} method defined in \emph{Bridge1DLT};
    \item The transaction burns the tokens on 1DLT;
    \item The \emph{Bridge1DLT} emits a custom event with the address of the receiver and the amount;
    \item The \emph{Mediator} detects the event and retrieves the information;
    \item The \emph{Mediator} builds a transaction to call the \emph{withdraw} method defined in \emph{BridgeEth} with the event information as parameters;
    \item The \emph{Mediator} sends the transaction to Ethereum;
    \item Ethereum executes the transaction, which calls the \emph{withdraw} method of \emph{BridgeEth};
    \item The method calls the \emph{transferFrom} defined in \emph{Token};
    % that transfers the tokens from the bridge address back to the user address.
 
\end{enumerate}

% story
%  List of the known risk bounded to bridges
%  List of the known reasons behind the major recent attacks - I can stress only on the studifity of users or bug in the code
%  List of things we have right now to provide safety
%   -- NO certification
%   -- We are not following already battle test solution (like optimism) which have suffered attacks
%   -- 
% \tbd{Potential risk of using bridges } 

As virtually all complex, interacting systems, bridges are exposed to security risks mainly related to:
\begin{itemize}
    \item {\bf Smart Contracts}: bugs in their code can be exploited for malicious behaviours;
    \item {\bf Underlying Blockchain}: the underlying blockchain can be hacked or behave in unexpected ways;
    % software failure, buggy code, human error, spam, and malicious attacks can possibly disrupt user operations
    \item {\bf User}: users not following best practices can incur non-secure behaviours;
    \item {\bf Censorship and Custodial}: bridge operators may act in malicious ways (e.g. they can suspend their activities or collude to gain sensitive information about the bridge’s users)\footnote{Applies to bridges that require the presence of trusted operators.}.
    \item {\bf Systematic financial}: the wrapped assets used by bridges to mint the canonical version of the original asset on a new chain, can be exploited, exposing the ecosystem to systematic risk.
    % bridges use wrapped assets to mint canonical versions of the original asset on a new chain. This exposes the ecosystem to systematic risk, as we have seen wrapped versions of tokens exploited.
    % Open issues – Given that bridges are in the nascent stages of development, there are many unanswered questions related to how bridges will perform in different market conditions, like times of network congestion and during unforeseen events such as network-level attacks or state rollbacks. This uncertainty poses certain risks, the degree of which is still unknown.
    
\end{itemize}

% Sources
% -solana https://www.euronews.com/next/2022/02/04/second-biggest-crypto-hack-attacks-wormhole-bridge-can-we-trust-defi-blockchain-technology
% https://rekt.news/polynetwork-rekt/           https://rekt.news/ronin-rekt/

There are numerous examples of bridge attacks that resulted in multimillion dollar losses.~\cite{rekt}.

% It is noteworthy to mention how a bridge hack happens from a vulnerability identified and exploited within the bridge contract. 
% As an example, the Wormhole attack \cite{wormholehack}, where a bug in the contract 
It is noteworthy to mention how a bridge hack mainly happens from a vulnerability identified and exploited within the bridge contract, such as the Wormhole attack \cite{wormholehack} or the Optimism smart contract bug \cite{optiSCbug}.
% or due to a user mistake (e.g., Optimism Wintermute case \cite{optiattack}). 
In the remaining cases, user mistakes take place,
% Other reason usually involve user mistakes,
such as in the  Optimism Wintermute case \cite{optiattack}, where a Wintermute user inserted the wrong destination address for a transaction.

% How we overcome it
% 
To overcome some of the previously mentioned risks, 
we give the user the ability to set up her/his 1DLT bridge without us taking control of the bridge or custody of the assets.
We offer reliable smart contracts that adhere to the community-tested security best practices, such as Optimism \cite{OptimisBridgeImplementation} or Polygon\cite{polygonBridge}).
Thus, an internal and external auditing procedure of the smart contracts, bridge,  and the node is conducted to ensure the users' safety.
For the external audit process, we use well-known auditors, such as CertiK\cite{certik}, Hacken\cite{hacken}, and Trail of Bits \cite{TrailofBits}.
As part of the internal audit process, we use a smart contract bytecode verification similar to the one of Etherscan \cite{etherscanVerification} and Sourcify~\cite{sourcify}.
To this end, to have the smart contracts verified, a user must share with us the transaction hash of the exploited smart contracts via a dedicated page. 

% We enable the smart contracts to operate only if the verification is successful.

%  A user deploying a bridge must share with us the addresses of the exploited smart contracts via a dedicated page. 
% The smart contracts are enabled only if the verification process is successful.
% To this end, a flag inside the smart contract is set only from an address under our control.

% Only after a successful verification are the smart contracts enabled. A flag inside the smart contract is set only from an address that we control in order to achieve this.

% Theft of user funds
% Liveness Issues
% Malicious Exploits
% Single Points of Failure/Centralization
% Poor Liquidity
% Technical Vulnerabilities
% Risk of Censorship

\section{ Experiments and Performance Discussion} \label{sec:experiments}
A set of preliminary experiments were performed to benchmark and test 1DLT. We remark that such experiments are only preliminary and do not fully account for the complexity of the experimental landscape of 1DLT (future versions of the present white paper will include a comprehensive experimental benchmark). 
% As such, we omit experiments to measure throughput, since those can be inaccurate without a proper benchmark.
% Instead,
We run experiments according to the following metrics:

\begin{enumerate}
\item {\bf Total transaction cost}: the cost of sending a transaction, which is computed as: 
% $$Cost_{total} = cost_{transaction} + fee_{gas} + fee_{DLT}$$ 
\begin{equation}
Cost_{total} = cost_{transaction} + fee_{gas} + fee_{DLT}
\end{equation}
where: $cost_{transaction}$ is the cost associated to the transaction, which may involve the execution cost of a smart contract or an amount of tokens; $ fee_{gas}$  and $fee_{DLT}$ are the fee costs associated to the transaction from the node and for the target DLT. It is important to note that the data field is where the difference between a transaction and an interaction lies; in a transaction, the field is empty; in an interaction, it has a value.
% trnasacione normale uguale a transazione di interazione la differenza è il campo data, 0 in una normale con quello che metti come contunto effettivo

\item {\bf Transaction finality}: the amount of time a user has to wait, on average, to obtain a confirmation of a transaction, generally measured in seconds.
In the case of 1DLT, the finality time includes the transaction processing time and consensus finality time of the public DLT.

\item {\bf Total smart contract deployment cost}: the cost of deploying a smart contract, which is computed as: 
% $$Cost_{total} = cost_{transaction} + fee_{gas} + fee_{DLT} + cost_{createContract} + cost_{data}$$
\begin{equation}
Cost_{total} = cost_{transaction} + fee_{gas} + fee_{DLT} + cost_{createContract} + cost_{data}
\end{equation}
where: $cost_{transaction}$ is the cost associated with the contract creation transaction; $fee_{gas}$  and $fee_{DLT}$ are the fee costs associated to the transaction from the node and for the target DLT; $cost_{Createcontract}$ is the cost associated to a contract creation, fixed to 32000 gas; and $cost_{data}$ is the cost associate to the contract complexity.
% How long is the $cost_{data}$ field, which means how complex is the smart contract.
As of today, we do not support only the legacy format (before EIP 2930~\cite{eip2930}) 
 
% 1559 inizio cambiamento, formato del cambiamento del gas
% 2718 introduce la nozione definizione del nuovo tipo di transazione tipo 0 legacy e 1 EIP 2930
% 2930 introduzione effettiva

% We note that right now we support only the legacy transaction type the EIP will be supported laterWe refer to the whipaper to the calculation \cite{} 

\item {\bf Throughput}: the transaction rate of a blockchain, measured in transactions per second (TPS). It is known that throughput is not the inverse of latency. For example, the transaction throughput for Bitcoin is about $7 tx/second$~\cite{bitcoinFinality} due to relatively small blocks and long block time. Instead, Ethereum has a short block time but tiny blocks, which results in a $15 tx/second$~\cite{Ethereumfinality}. 1DLT's  throughput is limited by the total throughput of public blockchains that CaaS connects to. In fact, all transactions submitted to CaaS by the QPQ Ethereum nodes are forwarded to the public blockchains like Hedera and Algorand. Therefore, 1DLT  throughput increases proportionally with the throughput of the blockchain CaaS connects to.

\end{enumerate}
The experiments are executed on an Azure Virtual Machine\footnote{https://azure.microsoft.com/} configured as $Standard\_D2\_v3$\footnote{https://docs.microsoft.com/en-gb/azure/virtual-machine/dv3-dsv3-series}, with 2 vCPUs, 8 GB of RAM, 256 GB SSD, and running \emph{Ubuntu 21.10}. We use Hedera Consensus Service (HCS)~\cite{HCS} on the Testnet as the consensus resource. 
We simulate the Web3 API interaction using Web3.js API~\cite{web3APIjs}.
We use Metamask~\cite{metamask} as the wallet application to verify the state of the transactions.

%We support mainstream development environments, such as Hardhat~\cite{devEnv}, Truffle~\cite{truffle} and REMIX IDE~\cite{remix}. 
The experiments have been developed using the Hardhat development environment~\cite{hardhat}, as it is the de-facto standard tool for developing dapps~\cite{solidityReport}.

%###### EXPERIMENTS

\begin{figure}[!htbp]
\centering
\includegraphics[width=0.34\textwidth, keepaspectratio]{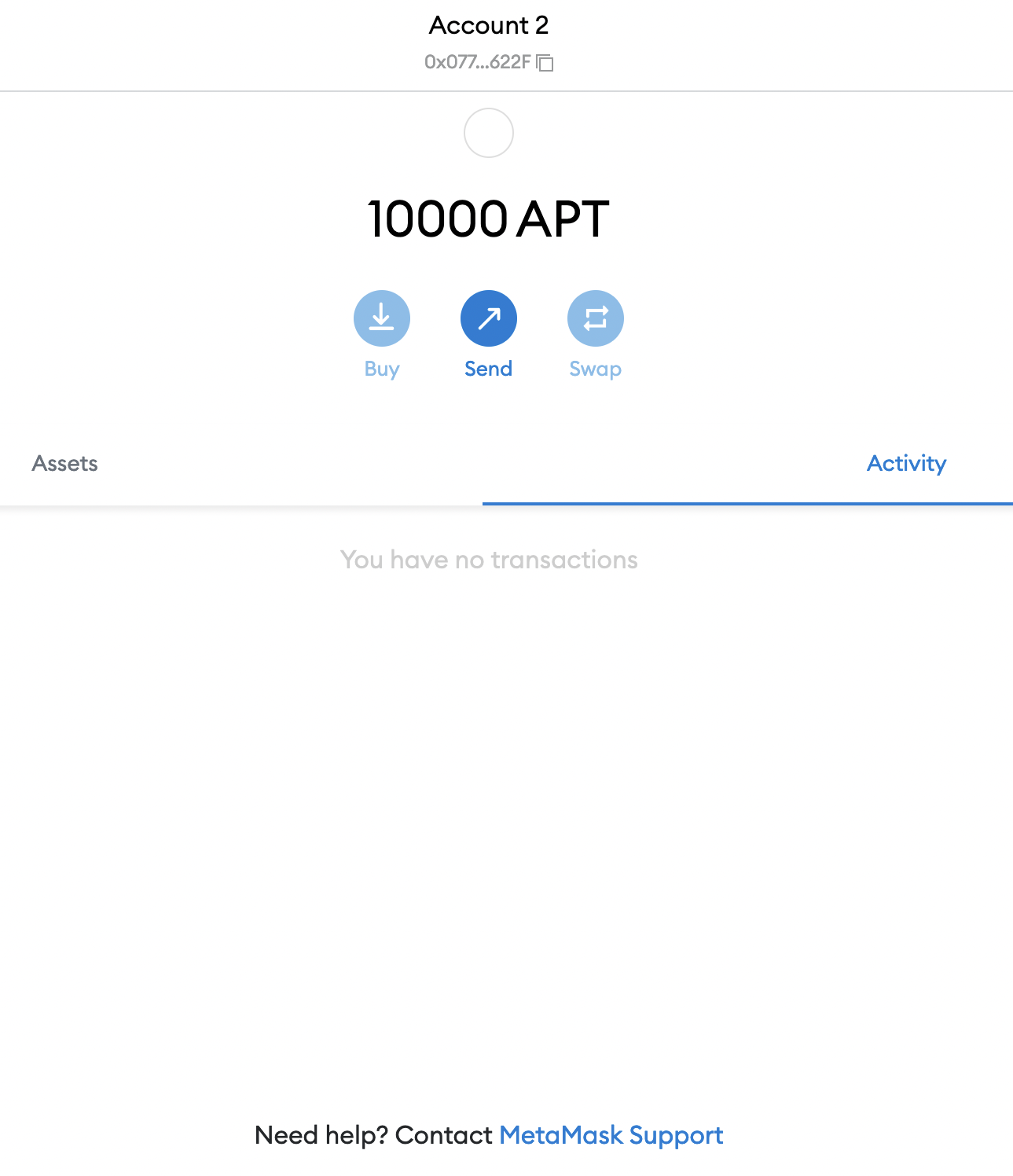}
\includegraphics[width=0.24\textwidth, keepaspectratio]{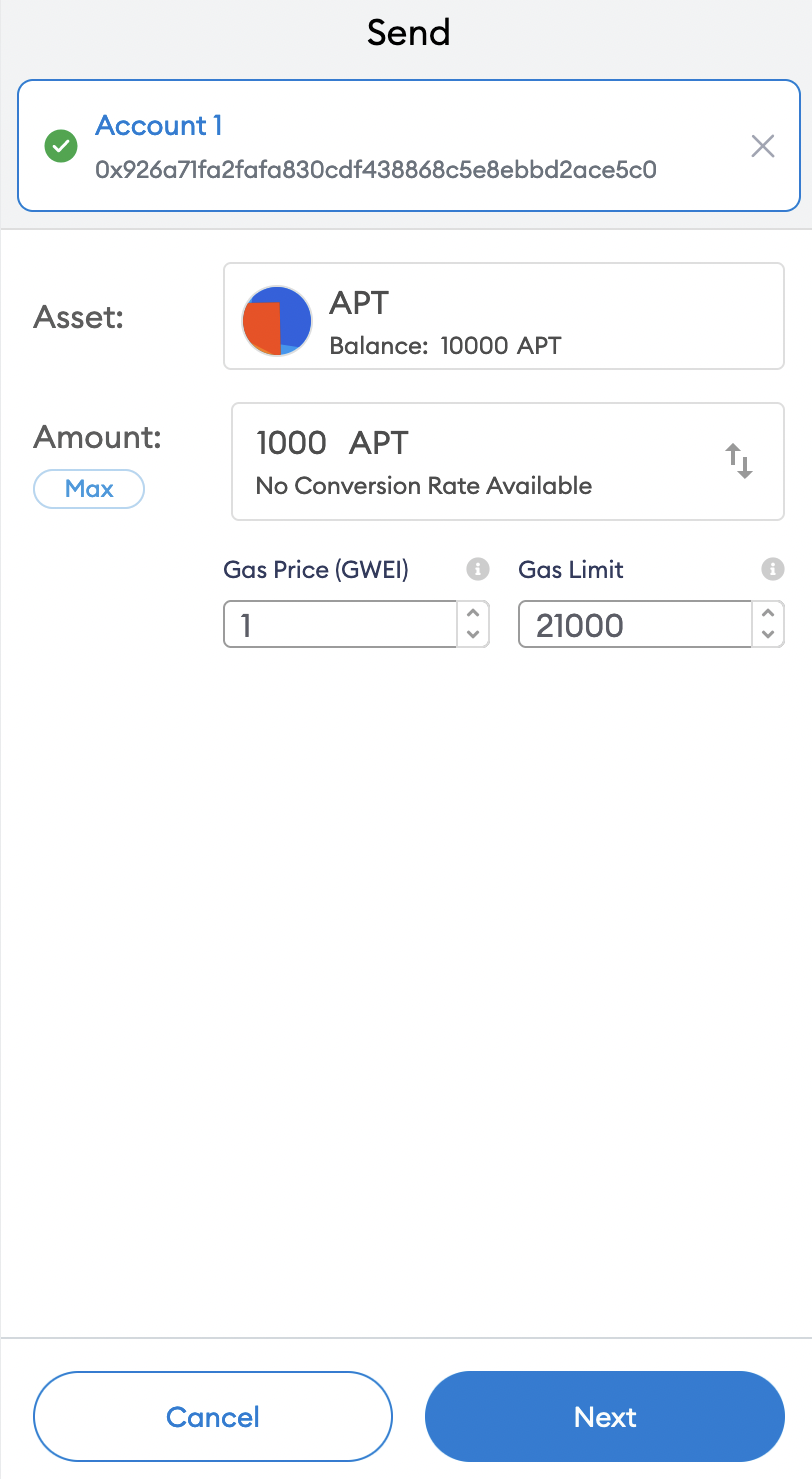}
\includegraphics[width=0.24\textwidth, keepaspectratio]{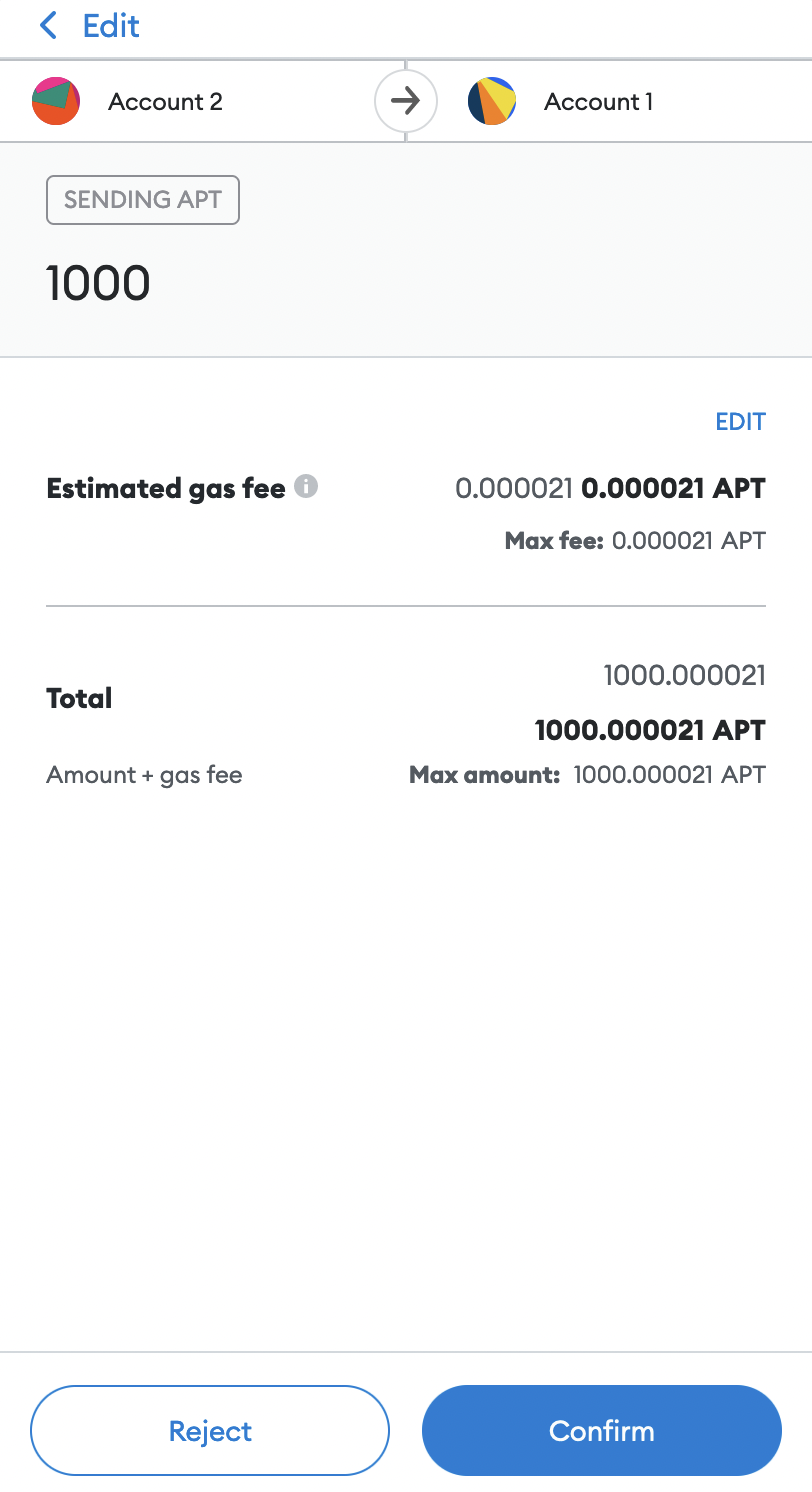}
\caption{Setup for Send transaction from Alice to Bob  }\label{fig:exp1}
\end{figure}

\begin{figure}[!htbp]
\centering
\includegraphics[width=0.24\textwidth, keepaspectratio]{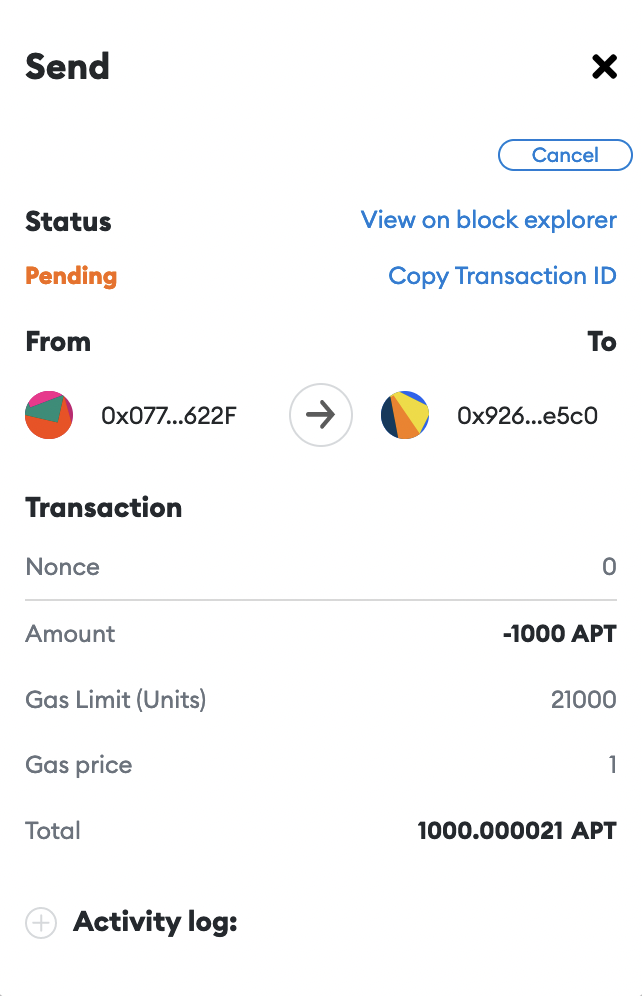}
\includegraphics[width=0.24\textwidth, keepaspectratio]{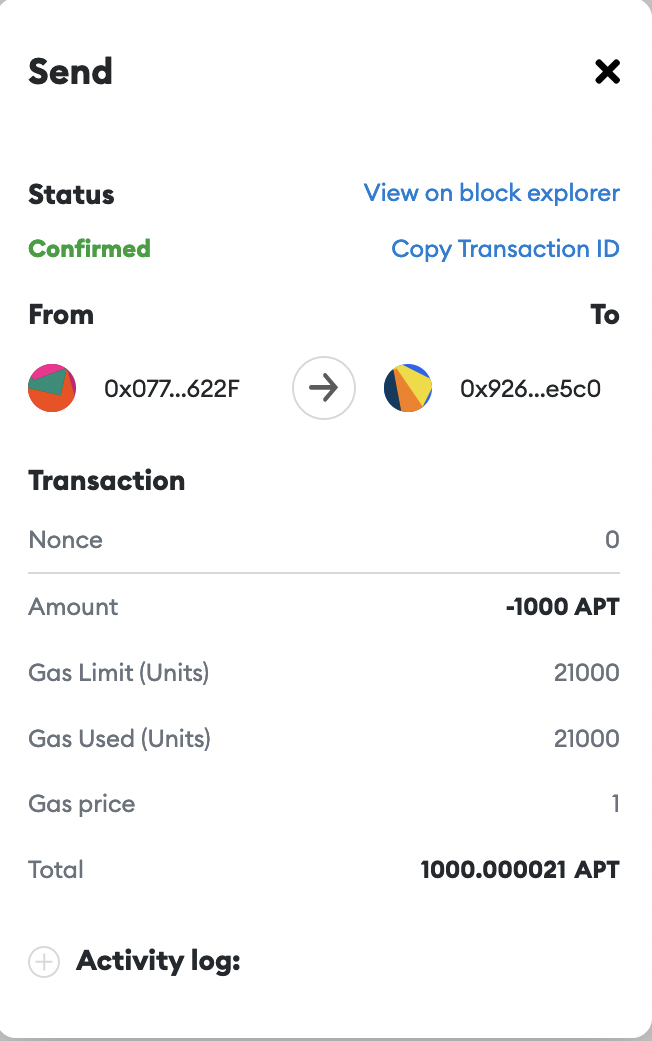}
\includegraphics[width=0.34\textwidth, keepaspectratio]{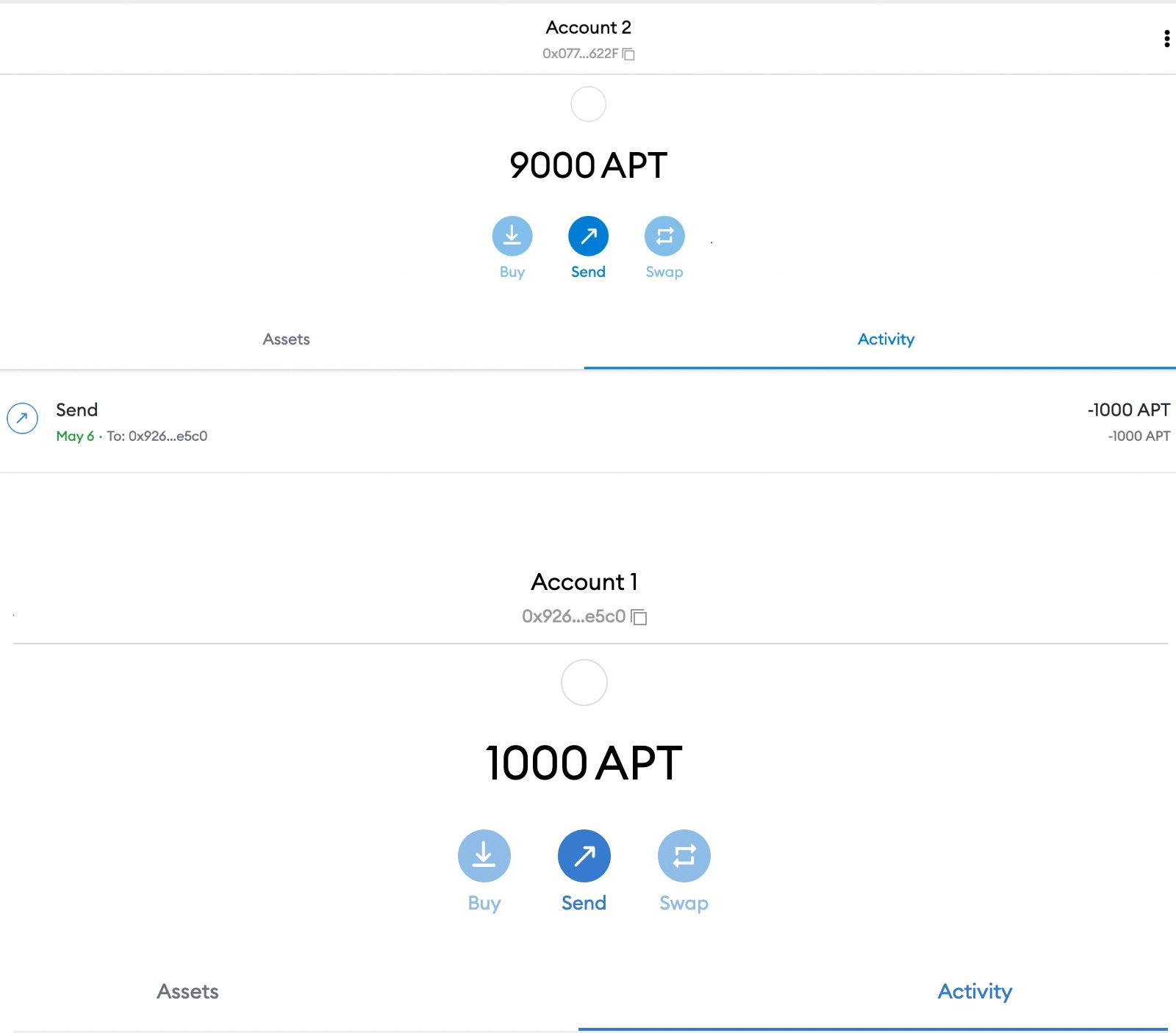}
\caption{Send transaction from Alice to Bob state transition and account update   }\label{fig:exp2}
\end{figure}

\paragraph{\textnormal{\textit{Total transaction cost:}}} 
We consider the token in Alice's 1DLT network, with token name and symbol $Alice\_Token$ and APT, respectively.
In Figure \ref{fig:exp1}, we show the steps done from Metamask's user interface to transfer tokens from Alice to Bob: first, we specify Bob's account as the destination, the APT token as the asset, and 1,000 as the amount. Second, we check the calculated fees and send the transaction. Initially, the transaction state is on pending, then, after 6 seconds, the state changes from pending to confirmed, allowing the balance update for Alice and Bob (see Figure \ref{fig:exp2}).

The total cost ($Cost_{total}$) for Alice is as follows: $fee_{Gas}$ is $0.000021$ $Alice\_Token$, $cost_{transaction}$ is 1000 $Alice\_Token$, and  $fee_{DLT}$ is 0.00051779 \emph{HBAR}, which is 0.00000003 $Alice\_Token$ (assuming that  1 ETH = 1 $Alice\_Token$).
As such, the cumulative cost will be $1,000.00002103$ $Alice\_Token$ and the cumulative cost for the fees is $0.00002103$ (0.056 USD). 
On the Ethereum Testnet (Ropsten \cite{ethTestnets}) the cumulative cost is $0.00005093$ (total of 0.14 USD), while on
the Ethereum mainnet, with a gas fee of 47 $Gwei$, is  $0.000819$ (for a total of 2.95 USD).
% 
% sent transaction  

\paragraph{\textnormal{\textit{Transaction finality:}}}
We evaluate the consensus finality of 1DLT using a client app that generates payment transactions, and submits them to a QPQ Ethereum node in its 1DLT network. We compute the overall time from the generation of the transaction to the balance update in the Metamask wallet as:
\begin{equation}
Overall\_time = Generate\_Send_{tx}+ 1DLT\_Finality_{tx} + Update\_wallet_{tx} 
\end{equation}
% $$Overall\_time = Generate\_Send_{tx}+ 1DLT\_Finality_{tx} + Update\_wallet_{tx}  $$ 
where:
\begin{itemize}
    \item $Generate\_Send_{tx}$ is the time it takes our client application to generate and send the payment transaction to a QPQ Ethereum Node of 1DLT;
    \item $1DLT\_Finality_{tx}$ is the time for 1DLT to process a transaction;
    \item $Update\_wallet_{tx}$ is the time it takes Metamask wallet to update the balance via a call sent by 1DLT.
\end{itemize}

\begin{figure}[!htbp]
\centering
\includegraphics[width=\textwidth, keepaspectratio]{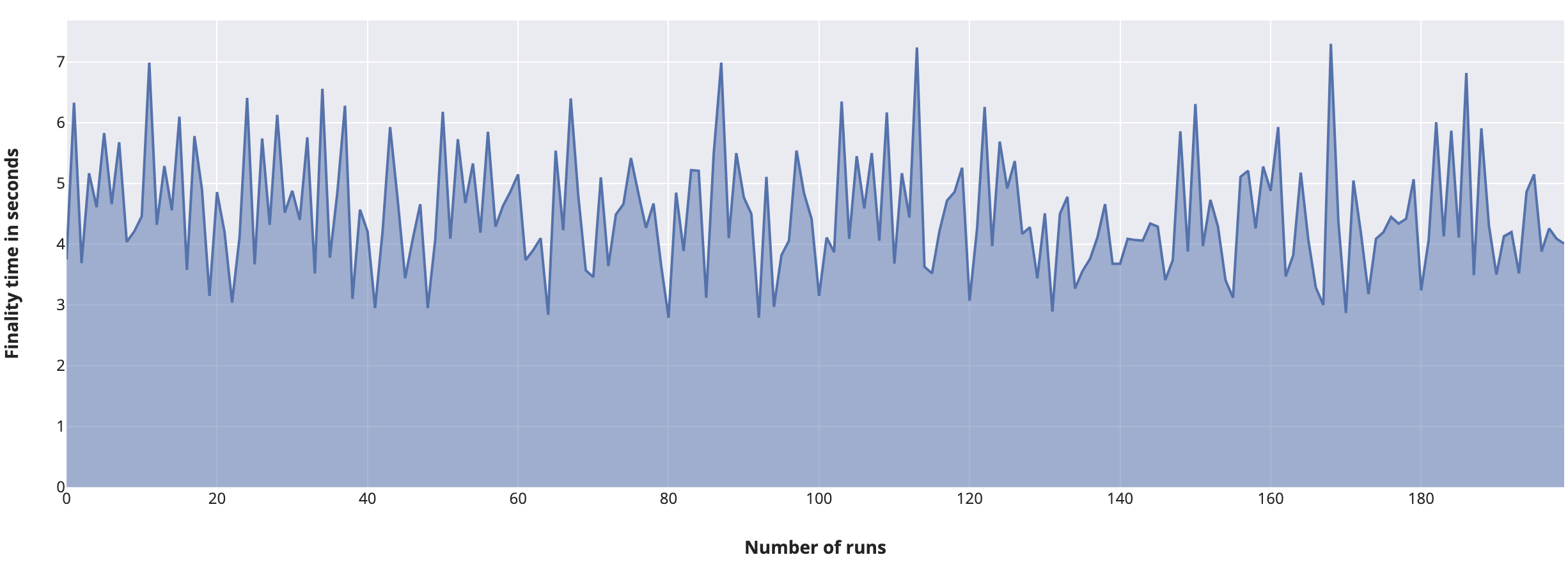}
\caption{Transaction Finality experiment Diagram }\label{fig:finalityGraph}
\end{figure}

In Figure \ref{fig:finalityGraph}, we present the experiment results, showing the $Overall\_Finality\_Time$ that we measured executing 200 transactions. We observe that average execution time for $Overall\_Finality\_Time$ is \textit{4.526} seconds.
%Where, the total time spend for $Generate\_Send_{tx}$ + $Update\_wallet_{tx}$ is around 400 milliseconds, i.e. 0.4 seconds.

\paragraph{\textnormal{\textit{Smart contract deployment cost:}}} 
% what we did
% which environment we used
% for each evn or at least one
%  - step to do the deploy
%  - show the ooutput/result of the deploy
% 
% Generally, to develop a Dapp you need to do both back end  and frontend deployemnt .With our solution you still need to build the frontend, 
We consider the example in \ref{ex:1} where Alice deploys the smart contract for her NFTs Auction dapp.
%and her costumers (i.e., Bob) interact with it.
% 
To deploy the auction smart contract, we write a deployment script in JavaScript. We add the 1DLT node information,  Node IP address, chain ID and private key of Alice account to the Hardhat configuration file, \emph{hardhat.config.js}.
Then, from terminal, we execute the deployment script with the command: 
% 
% We write the deployment script (\ref{Appendix} deploySmartContract.js) to deploy the auction smart contact (\ref{Appendix} NFT\_Auction smart contract)
% Then, we set 1DLT as the endpoint for the communication by adding the references of the 1DLT node in the Hardhat configuration (hardhat.config.ts) file, which are Node ip address, chainID and the private key of your account. (see \ref{} code snippet). 
% 

\begin{lstlisting}[language=bash,basicstyle=\ttfamily\scriptsize]
  $ pnpm hardhat run --network AliceNetwork deploySmartContract.js
\end{lstlisting}
Once the deployment is completed, we receive the address of the created smart contract, like:
\begin{lstlisting}[language=bash,basicstyle=\ttfamily\scriptsize]
  $ Contract deployed to address: 0x6cd7d44516a20882cEa2DE9f205bF401c0d23570
\end{lstlisting}

The transaction cost to deploy the smart contract on 1DLT is $0.000013402$ Alice token (suppose that  1 ETH = 1 $Alice\_Token$).
On the Ethereum Testnet (Ropsten) the cumulative cost is $0.0015402$ (total of 1.74 USD), while on the Ethereum mainnet, with a gas fee of 30 $Gwei$, is  $0.0117055$ (for a total of 13.91 USD). We note that the cost to interact with a smart contract, that is, to call a method that changes the state (e.g., a set method), is calculated the same as a transaction.
In fact, under the hood, a setter method is implemented by sending a transaction.

\paragraph{\textnormal{\textit{Transaction per second (TPS): }}} 

We evaluate the performance of CaaS using Hedera as the consensus resource.
We use a different Azure virtual machine configuration than before. We run CaaS on an Azure VM in the Switzerland North region configured as Standard DS3 v2, with 4 vCPUs, 14 GB of RAM, 1 TB SSD, and running Ubuntu 20.04. 
% Then, we consider two configuration scenarios to simulate the clients: (i) a client is hosted on a VM configured with Standard D2s v3, 2 vCPUs, 8 GiB memory, and running Ubuntu 20.04.
%(ii) a client is a MacBook Pro 2,3 GHz 8-Core Intel Core i9 and 32 GB 2667 MHz DDR4 with a bandwidth ranging from 10 to 13 1mbps for download, and from 5 to 55 mbps.
% 
Then, to simulate the client we use a VM configured with Standard D2s v3, 2 vCPUs, 8 GiB memory, and running Ubuntu 20.04.
We want to underline that in this experiment, we configured CaaS to run on a small VM to prove that it is very lightweight and can achieve high performance even with this setup. Ideally, and in the production environment, CaaS will run behind a Kubernetes cluster that allows it to scale up with the increased transaction requests.

The client runs a Python v3.8.10 script that generates a total of ten thousand transactions across five Hedera topics (which corresponds to five communication channels in CaaS), sends transactions to CaaS, and waits for confirmations from CaaS.

% Thus, each client is using  5 unique channels, for a total of 20 channels.

Running the experiment 10 times with a single client results in an average of 1120 tps.
% We run the experiment 10 times with 1 client, obtaining an average of, 1120 tps.

%underline the capability of single node making more than 1k TX. 
The results are promising, as this experiment proves that a single client can process around 1120 transactions per second with a minimal CaaS setup as discussed above.

\subsection{Discussion} \label{sec:FCEdiscussion}
% \section{1DLT as full-fledged computational environment } \label{sec:FCE}
% // why 1DLT is not Raspberry,
% \paragraph{\textnormal{\textit{Experiment 4: Energy usage, Computational cost and power.}}}
% \paragraph{\textnormal{\textit{Experiment 4: 1DLT as full-fledged computational environment.}}}
% vce 
% full fledge vitrtual cumputation enviroment

% intro
% Web3, in the context of Ethereum, refers to decentralized apps that run on the blockchain.
We now discuss the energy consumption and costs of the Ethereum network, arguing the that 1DLT offers a better trade-off between costs and performances.
Ethereum currently uses a proof-of-work~\cite{PoW} consensus mechanism to ensure security and decentralisation.
To this end, Ethereum requires massive computational power to run, with high operating costs and energy consumption~\cite{EthEnergyConsumption}.
% Ethereum currently requires a massive computational power to run, with high operating costs and energy consumption~\cite{EthEnergyConsumption}.
% To ensure security and decentralisation, Ethereum uses the proof-of-work~\cite{PoW} consensus mechanism.
% 
%problems
% high energy consumption
In a PoW-based setting, transactions are confirmed by miners, which can add a new block to the ledger only after solving an algorithmic puzzle that entails an associated computational cost. 
%The required high computation is in the form of a cryptographic hash puzzle, exploited to keep the network secure, that a miner has to solve faster than any other miner.
% Therefore, the required cost to procure a block increase exponentially with the puzzle difficulty and obliges the miner to invest in powerful hardware, creating a race to power-hungry mining equipment.
In particular, miners compete against each others for the creation of each new block.
This competition forces miners to invest in increasingly powerful hardware, creating a race to energy-hungry mining equipment.
% In proof-of-work, transaction confirmation is performed by miners, which are eligible to add a new block only with an associated computational cost.
% The cost is in the form of a cryptographic hash puzzle that a miner has to solve faster than any other miner, to have its block accepted to the main chain and claim the reward.
% The puzzle consists in finding a nonce value, which hashed together with the block contents yields a value lower than a specified threshold, referred to as the 'difficulty'.
% % 
% The speed at which a nonce is found is directly proportional to the hashing power in the network.
% In turn, this relates to the number of miners, as more miners typically bring more computational power.
% More specifically, the higher the network hashing power, the faster a nonce is found for the new block.
% % 
% To keep the block production at a fixed average rate, the puzzle difficulty is adjusted accordingly:
% when the hashing power increases, so does the mining difficulty.
% Therefore, while mining competition keeps the network secure, it also increases the energy consumption required to produce a block.
% 
% The competition also obliges the miner to invest in powerful hardware, creating a race to power-hungry mining equipment.

\begin{figure}[!htbp]
\centering
\includegraphics[width=0.60\textwidth, keepaspectratio]{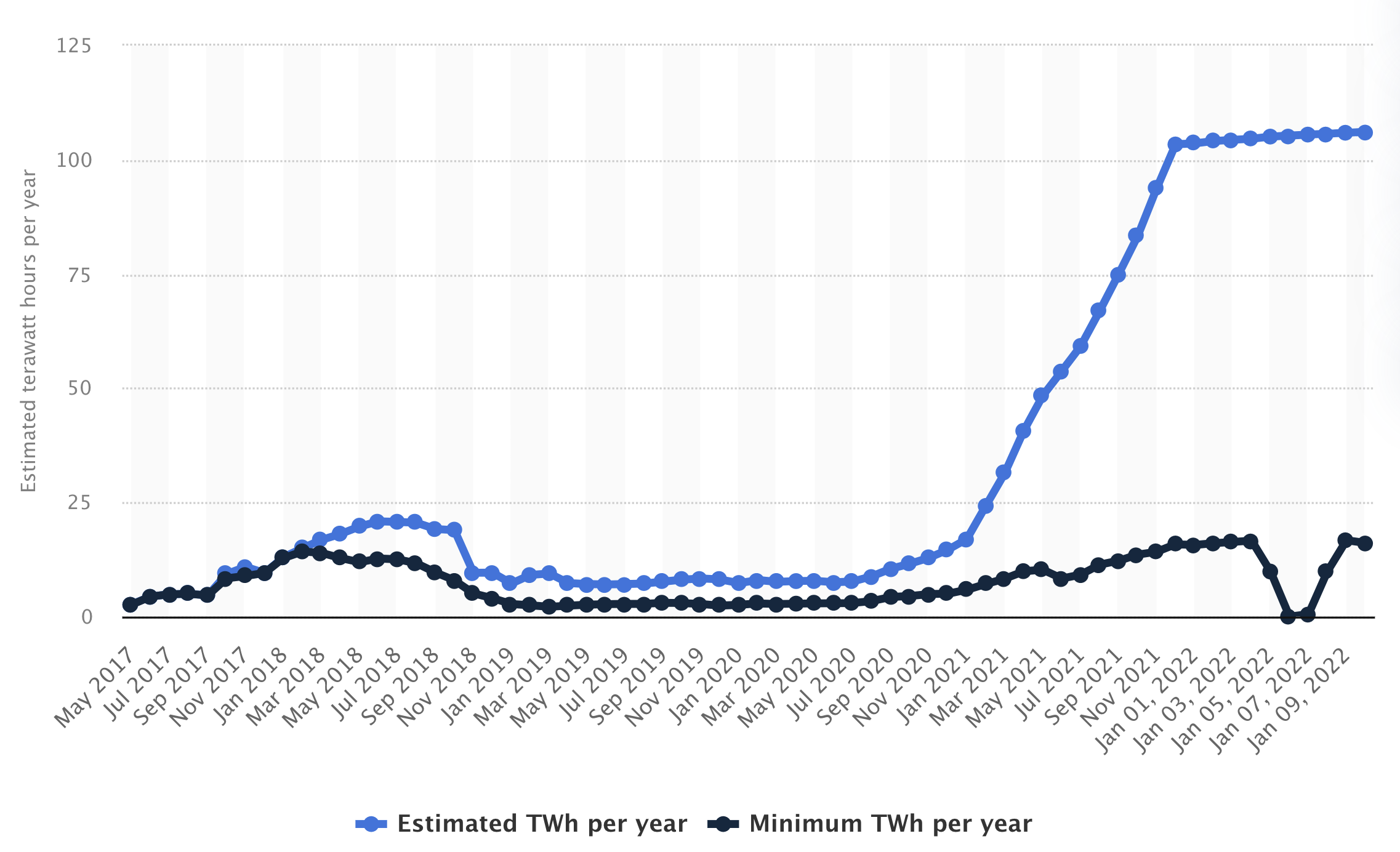}
\caption{ Worldwide Ethereum energy consumption  
% ~\cite{ethEnergy}
}\label{fig:EthereumWattConsumption}
\end{figure}

As of now, the current Ethereum energy consumption is above 100 TWh/year~\cite{ethEnergy, ethFootprint} (see Figure \ref{fig:EthereumWattConsumption}), which is roughly the energy consumption of a country like Austria.
Also, the annual carbon footprint is 53.81 Mt CO~\cite{ethFootprint}, which is around the carbon footprint of Singapore.
The average footprint for one transaction is 227.67 kWh~\cite{ethFootprint}, which is slightly 100 kWh (which in turn is more than the footprint of 100,000 VISA transactions~\cite{visa}).
% 

% All below is Rephrased from sources here
% https://blog.lamden.io/proof-of-work-kills-the-earth-e687d3e83ec9#:~:text=Multiply%20the%20total%20number%20of,PFLOPS%20(yes%2C%20petaFLOPS).
% \tbd{Computation cost} 
% At the same time, the Ethereum network has an hash rate of 884,510 GH/s~\cite{hashratechart}. 
% % oold 1126674x15 = 16,900,110
% Considering that a block needs 15 seconds to be finalized~\cite{ethstat}, 
% a miner must wait for 13,267,650 GH to gain the block's reward.
% %
% Thus, if we consider 10,000 miners that hash at five hashes per second, the network needs 100,000 hashes before finding a solution. 
On the other side, if we take into account 10,000 miners that hash at five hashes per second, the network needs 100,000 hashes before finding a solution. 
Due to the PoW mechanism, only the 0.001\% hashes are successful.
Currently, the Ethereum network has a hash rate of 884,510 GH/s~\cite{hashratechart} and a block needs 15 seconds to be finalized~\cite{ethstat}, meaning that a miner must wait for 13,267,650 GH to gain the block's reward.
Thus, the Ethereum network is currently wasting 99.999\% of its computational power hashing random numbers.

% operating cost
Each finalized block includes 30 million Gas, which is the amount of Gas used for all the transactions in a block~\cite{gas}.
% todo: check numbers below and add current date -----done
The current transaction fees for 30 million of consumed gas is more than 1 Ether, which (as of July 2022) is valued at 1,479 USD.
This implies that the computation costs of the Ethereum Network are around 133 USD per second, which is $\sim$25 times more than 15 days of an EC2 instance (currently around 20 USD).

So far, we have discussed the computational power that is needed for keeping the Ethereum network live and running. 
% Now, in order to grasp the computational power of the Ethereum network itself, following \cite{web3fraud}, we consider a very basic computational task like adding two 256-bit integers.
% This operation costs 3 Gas~\cite{opcodes}. Thus, as the total compute of the Ethereum network is 2 million gas/second, the Ethereum Network may perform 600,000 additions per second.
Now, in order to estimate the cost of the Ethereum network itself (measured in gas), following \cite{web3fraud}, we consider a very basic computational task like adding two 256-bit integers.
Because this operation costs 3 Gas~\cite{opcodes} and the Ethereum network's total compute is 2 million gas/second, the Ethereum network may perform 600,000 additions per second.
% This operation costs 3 Gas~\cite{opcodes}. Thus, as the total compute of the Ethereum network is 2 million gas/second, the Ethereum Network may perform 600,000 additions per second.
% and as a new block occurs every 15 seconds, the total compute of the Ethereum network as 2 million gas/second,
%
In comparison, Raspberry Pi 4~\cite{raspberry}, a 45 USD single-board computer with four processors running at 1.5 GHz, can perform around 3,000,000,000 additions per second.
As such, the Ethereum network, considered as a general-purpose computational environment, has roughly 1/5,000 of the computing power of a Raspberry Pi 4.
At the current gas price, this means that performing 256-bit additions on the Ethereum network, costs about 60 USD per month.
% 60$ al mese circa

%contents 
%  - no PoW 
%  - external consensus 
%  - low energy usage
%  - low computation cost
%  - high tps  ?
%  - little time to finality ?

% \tbd{Closing}
% sources https://hedera.com/blog/going-carbon-negative-at-hedera-hashgraph
% https://hedera.com/ucl-blockchain-energy
% https://medium.com/@ismailvohra/hedera-hashgraph-greener-alternative-to-blockchain-3e6d2ac12b05

Ethereum as a computational environment has the drawback to be expensive and highly energy consumption. The energy consumption will be reduced with the introduction of the Ethereum merge~\cite{Themerge} (announced to be delivered in September 2022) as the consensus mechanism will shift from PoW to PoS. However, this change does not impact the overall price reduction, as the gas price will not be affected.

As mentioned at the beginning of this discussion, 1DLT follows a modular approach and separates the EVM-based computational layer from the consensus layer, minimising energy consumption and computational effort.
Thanks to the deployment of CaaS, the consensus engine does not require a mining algorithm in the consensus retrieval.
Additionally, 1DLT offers the same level of computational power as Ethereum at a significantly lower cost, as shown in experiments 1 and 3 in Section \ref{sec:experiments}.
As an example, the cost to execute 50,000 transactions is 0.04 USD.
% 
% 
%In the end, we aim to enable users to deploy the whole infrastructure, not only the transactions, enabling 1DLT to be a full computational environment.

% 

% 1DLT minimize the energy usage and computational waste as as no mining algorithm is involved in the consensus retrieval.  
% In 1DLT the energy usage and computational waste are minimized as no mining algorithm is involved in the consensus retrieval.
% Instead for the computational power it can perform an higher number of opcodes thanks block confirmation rate. 

% Also, it provides a full computational environment, meaning that the user buys the entire infrastructure and a finite set of transaction.
% Also, it provides a full computational environment, meaning that the user buys the entire infrastructure and a finite set of transaction.

% 1DLT reduces energy consumption and computational cost, improving computational power without sacrificing security and decentralisation.
% To this end, it leverages an external consensus source (CaaS) for consensus retrieval, which brings two benefits. First, our solution can rely on other proof rather than only PoW (e.g., Hedera Proof-of-Stake), meaning a reduced energy consumption as there is no need for miners. Second, by leveraging on our tps, our solution can provide a higher computation power with less computation cost.
% Also, we do not waste power in hashing numbers as no mining is involved

\section{ Conclusion and Future works }\label{sec:conclusion}
Scaling solutions are crucial to increasing the Ethereum network's capacity in terms of speed and throughput, but they come at the cost of reduced decentralisation, increased transaction finality times, or loss of trustlessness.\\
\\
1DLT, inspired by the user experience of Cloud Service Providers (CSP) and Web-based applications, overcomes the limitations of the existing scaling solutions enabling low gas fees, high transaction throughput, and fast transaction finality. Additionally, 1DLT removes the risk associated with the L2 governance and fraud detection, since all the transactions processed in 1DLT networks are submitted to the consensus protocols of L1 public DLTs. Lastly, the programmability and user experience of the Ethereum ecosystem is maintained thanks to an EVM-based architecture.\\
\\
We demonstrated the feasibility of our architecture with a set of preliminary experiments in Section \ref{sec:experiments}, benchmarking transaction costs and finality. \\
\\
Future work includes: (i) a proper experimental benchmark to execute advanced experiments that fully accounts for the real-world landscape of L2 solutions and 1DLT; 
(ii) enhance the bridge with support for blockchains that are not compatible with EVM;
% increase interoperability between 1DLT and other DLT through bridges to enable the exchange of assets; 
(iii) the integration of 1DLT with Trusted Execution Environments; (iv) full integration in the QPQ Atomic Swap Engine (ASE), and with a QPQ proprietary wallet that will enhance the Metamask solution; (v) provision of user tools like Etherscan~\cite{etherscan} for Ethereum or DragonGlass~\cite{dragonglass} for Hedera, to audit the status of the blockchains in the 1DLT ecosystem.

\bibliographystyle{ieeetr}
\bibliography{Main}

\begin{thebibliography}{10}

\bibitem{wood2014ethereum}
G.~Wood, ``Ethereum: A secure decentralised generalised transaction ledger.,''
  {\em Ethereum Project Yellow Paper}, 2014.

\bibitem{ethtps}
``{Ethereum TPS}.'' \url{https://ethtps.info/}, 2022.
\newblock [Online].

\bibitem{apes}
``{Bored ape crush Ethereum}.''
  \url{https://www.cnet.com/personal-finance/crypto/bored-ape-yacht-club-just-broke-the-ethereum-blockchain/},
  2022.
\newblock [Online].

\bibitem{EthEnergyConsumption}
``{Ethereum energy consumption}.''
  \url{https://ethereum.org/en/energy-consumption/}, 2022.
\newblock [Online].

\bibitem{ethEnergy}
``{Ethereum energy consumption statistics}.''
  \url{https://www.statista.com/statistics/1265897/worldwide-ethereum-energy-consumption/},
  2022.
\newblock [Online].

\bibitem{web3fraud}
N.~Weaver, ``{The Web3 Fraud}.''
  \url{https://www.usenix.org/publications/loginonline/web3-fraud}, 2021.
\newblock [Online].

\bibitem{L2eth}
``{Ethereum Scaling}.'' \url{https://ethereum.org/en/developers/docs/scaling/},
  2022.
\newblock [Online].

\bibitem{optrollup}
Optimism, ``{Optimistic Rollups}.''
  \url{https://ethereum.org/en/developers/docs/scaling/optimistic-rollups/},
  2022.
\newblock [Online].

\bibitem{zkrollups}
``{Zero-Knowledge Rollups}.''
  \url{https://ethereum.org/en/developers/docs/scaling/zk-rollups}, 2022.
\newblock [Online].

\bibitem{statechannels}
``{State Channels}.''
  \url{https://ethereum.org/en/developers/docs/scaling/state-channels/}, 2022.
\newblock [Online].

\bibitem{sidechains}
``{Sidechains}.''
  \url{https://ethereum.org/en/developers/docs/scaling/sidechains/}, 2022.
\newblock [Online].

\bibitem{plasma}
``{Plasma}.'' \url{https://ethereum.org/en/developers/docs/scaling/plasma/},
  2022.
\newblock [Online].

\bibitem{validium}
``{Validium}.''
  \url{https://ethereum.org/en/developers/docs/scaling/validium/}, 2022.
\newblock [Online].

\bibitem{StarkwareCairo}
Starkware, ``{Starkware Cairo}.'' \url{https://starkware.co/cairo/}, 2022.
\newblock [Online].

\bibitem{Starkware}
Starkware, ``{Starkware Libs}.'' \url{https://github.com/starkware-libs/},
  2022.
\newblock [Online].

\bibitem{ClientsNode}
``{Ethereum client and node definition}.''
  \url{https://ethereum.org/en/developers/docs/nodes-and-clients/}, 2022.
\newblock [Online].

\bibitem{Geth}
``{Geth}.'' \url{https://geth.ethereum.org/docs/}, 2022.
\newblock [Online].

\bibitem{Erigon}
``{Erigon}.'' \url{https://github.com/ledgerwatch/erigon}, 2022.
\newblock [Online].

\bibitem{Hederamirrorservice}
``{Hedera mirror service}.''
  \url{https://hedera.com/learning/hedera-hashgraph/what-is-the-hedera-mirror-network},
  2022.
\newblock [Online].

\bibitem{Trie}
``{Merkle Patricia Trie}.'' \url{https://eth.wiki/fundamentals/patricia-tree},
  2021.
\newblock [Online].

\bibitem{EVM}
``{EVMe}.'' \url{https://ethereum.org/en/developers/docs/evm/}, 2021.
\newblock [Online].

\bibitem{EVMone}
``{EVMone}.'' \url{https://github.com/ethereum/evmone}, 2021.
\newblock [Online].

\bibitem{metamask}
``{Metamask}.'' \url{https://metamask.io/}, 2021.
\newblock [Online].

\bibitem{hardhat}
``{Hardhat}.'' \url{https://hardhat.org/}, 2021.
\newblock [Online].

\bibitem{web3APIjs}
``{web3.js API}.'' \url{https://web3js.readthedocs.io/en/v1.7.4/}, 2021.
\newblock [Online].

\bibitem{leveldb}
``{Leveldb database}.'' \url{https://github.com/google/leveldb}, 2022.
\newblock [Online].

\bibitem{Sled}
``{Sled database}.'' \url{https://github.com/spacejam/sled}, 2022.
\newblock [Online].

\bibitem{Erigonstage}
``{Erigon stage sync}.''
  \url{https://github.com/ledgerwatch/erigon/blob/devel/eth/stagedsync/README.md},
  2022.
\newblock [Online].

\bibitem{ethBridgeIntro}
``{Introduction to blockchain bridges}.''
  \url{https://ethereum.org/en/bridges/}, 2022.
\newblock [Online].

\bibitem{ethBridgeIntro2}
``{Blockchain bridges}.''
  \url{https://ethereum.org/en/developers/docs/bridges/}, 2022.
\newblock [Online].

\bibitem{InteroperabilityTrilemma}
``{The interoperability trilemma}.''
  \url{https://blog.connext.network/the-interoperability-trilemma-657c2cf69f17},
  2022.
\newblock [Online].

\bibitem{bridgeClassification}
``{Blockchain bridges classification}.''
  \url{https://li.fi/knowledge-hub/bridge-classification/}, 2022.
\newblock [Online].

\bibitem{binancebridge}
``{Binance bridge}.'' \url{https://www.bnbchain.org/en/bridge}, 2022.
\newblock [Online].

\bibitem{connext}
``{Connext bridge}.'' \url{https://bridge.connext.network/}, 2022.
\newblock [Online].

\bibitem{hop}
``{Hop}.'' \url{https://app.hop.exchange/}, 2022.
\newblock [Online].

\bibitem{rekt}
``{Leaderboard of Ethereum bridge attacks}.''
  \url{https://rekt.news/leaderboard/}, 2022.
\newblock [Online].

\bibitem{wormholehack}
``{Wormhole hack}.'' \url{https://rekt.news/wormhole-rekt/}, 2022.
\newblock [Online].

\bibitem{optiSCbug}
``{Optimism smart contract bug}.''
  \url{https://cryptoslate.com/critical-bug-in-ethereum-l2\\-optimism-2m-bounty-paid/},
  2022.
\newblock [Online].

\bibitem{optiattack}
``{Optimism Attack}.''
  \url{https://cointelegraph.com/news/optimism-loses-20m-tokens-after-\\l1-and-l2-confusion-exploited},
  2022.
\newblock [Online].

\bibitem{OptimisBridgeImplementation}
``{The Optimism bridge}.''
  \url{https://ethereum.org/en/developers/tutorials/optimism-std-bridge-annotated-code/},
  2022.
\newblock [Online].

\bibitem{polygonBridge}
``{The Polygon-Ethereum Bridge}.''
  \url{https://docs.polygon.technology/docs/develop/ethereum-polygon/getting-started/},
  2022.
\newblock [Online].

\bibitem{certik}
``{CertiK}.'' \url{https://www.certik.com/}, 2022.
\newblock [Online].

\bibitem{hacken}
``{Hacken}.'' \url{https://hacken.io/}, 2022.
\newblock [Online].

\bibitem{TrailofBits}
``{Trail of Bits}.'' \url{https://www.trailofbits.com/}, 2022.
\newblock [Online].

\bibitem{etherscanVerification}
``{Etherscan smart contract verification}.''
  \url{https://etherscan.io/verifyContract}, 2022.
\newblock [Online].

\bibitem{sourcify}
``{Sourcify}.'' \url{https://docs.sourcify.dev/docs/intro}, 2022.
\newblock [Online].

\bibitem{eip2930}
``{EIPS 2930}.'' \url{https://eips.ethereum.org/EIPS/eip-2930}, 2022.
\newblock [Online].

\bibitem{bitcoinFinality}
``{Real time Bitcoin finality}.''
  \url{https://statoshi.info/d/000000006/transactions?viewPanel=6&orgId=1},
  2022.
\newblock [Online].

\bibitem{Ethereumfinality}
``{Real time Ethereum finality}.'' \url{https://ethtps.info/}, 2022.
\newblock [Online].

\bibitem{HCS}
``{Hedera Consensus Service}.'' \url{hedera.com/consensus-service}, 2021.
\newblock [Online].

\bibitem{solidityReport}
``{Solidity report 2021}.''
  \url{https://blog.soliditylang.org/2022/02/07/solidity-developer-survey-2021-results/},
  2022.
\newblock [Online].

\bibitem{ethTestnets}
``{Ethereum testnets}.''
  \url{https://ethereum.org/en/developers/docs/networks/}, 2022.
\newblock [Online].

\bibitem{PoW}
``{Ethereum Proof of work}.''
  \url{https://ethereum.org/en/developers/docs/consensus-mechanisms/pow/},
  2022.
\newblock [Online].

\bibitem{ethFootprint}
``{Ethereum energy footprints}.''
  \url{https://digiconomist.net/ethereum-energy-consumption/}, 2022.
\newblock [Online].

\bibitem{visa}
``{Ethereum energy consumption compared to VISA}.''
  \url{https://www.statista.com/statistics/1265891/ethereum-energy-consumption-transaction-comparison-visa/},
  2022.
\newblock [Online].

\bibitem{hashratechart}
``{Ethereum hashrate chart}.'' \url{https://etherscan.io/chart/hashrate}, 2022.
\newblock [Online].

\bibitem{ethstat}
``{Ethereum statistics}.'' \url{https://ethstats.net/}, 2022.
\newblock [Online].

\bibitem{gas}
``{Ethereum gas}.'' \url{https://ethereum.org/en/developers/docs/gas}, 2022.
\newblock [Online].

\bibitem{opcodes}
``{Ethereum opcodes}.''
  \url{https://ethereum.org/it/developers/docs/evm/opcodes/}, 2022.
\newblock [Online].

\bibitem{raspberry}
``{Raspberry datasheet}.''
  \url{https://datasheets.raspberrypi.com/rpi4/raspberry-pi-4-datasheet.pdf},
  2022.
\newblock [Online].

\bibitem{Themerge}
``{The Ethereum merge}.'' \url{https://ethereum.org/en/upgrades/merge/}, 2022.
\newblock [Online].

\bibitem{etherscan}
``{Etherscan}.'' \url{https://etherscan.io/}, 2021.
\newblock [Online].

\bibitem{dragonglass}
``{DragonGlass}.'' \url{https://app.dragonglass.me/}, 2021.
\newblock [Online].

\end{thebibliography}

\newpage
\noindent\textbf{Simone Bottoni}  is a Research Engineer at QPQ.  He is also a  PhD student in Computer Science at Insubria University, Varese, Italy, and holds an MSc in Computer Science from Insubria University, Varese, Italy.\\
\vspace{3pt}

\noindent\textbf{Anwitaman Datta} is a Senior Scientific Officer at QPQ. He is also an Associate Professor at the School of Computer Science and Engineering in NTU Singapore. He holds a PhD in Computer and Communication Sciences from EPFL, Switzerland, and a B. Tech in Electrical Engineering from IIT Kanpur, India.\\
\vspace{3pt}

\noindent\textbf{Federico Franzoni} is a Research Engineer at QPQ. He holds a PhD in Information Technology from the University of Pompeu Fabra, Barcelona, Spain, and an MSc in Computer Science from Sapienza University, Rome, Italy.\\
\vspace{3pt}

\noindent\textbf{Emanuele Ragnoli} is the CTO at QPQ. He holds a PhD in Systems Theory from the Hamilton Institute, NUI Maynooth, Ireland, and an MSc in Mathematics from Milan University, Italy.\\
\vspace{3pt}

\noindent\textbf{Roberto Ripamonti} is a Research Engineer at QPQ. He holds an MSc in Computer Science from Insubria University, Varese, Italy.\\
\vspace{3pt}

\noindent\textbf{Christian Rondanini} is a Research Engineer at QPQ. He holds a PhD in Computer Science from Insubria University, Varese, Italy, and an MSc in Computer Science from Insubria University, Varese, Italy.\\
\vspace{3pt}

\noindent\textbf{Gokhan Sagirlar} is a Research Engineer at QPQ. He holds a PhD in Computer Science from Insubria University, Varese, Italy, and a BSc in Computer Science from Ege University, İzmir, Turkey.\\
\vspace{3pt}

\noindent\textbf{Alberto Trombetta} is the CSO at QPQ. He is also  an Associate Professor of Computer Science at the Department of Theoretical and Applied Sciences of Insubria University. He holds a PhD in Computer Science from Turin University, Italy, and an MSc in Computer Science from Milan University, Italy.

\end{document}